\documentclass[a4paper]{article}

\usepackage[naustrian,american,british]{babel}	%geladene Sprachen; Aufruf durch z.B. "\selectlanguage{naustrian}"}
\usepackage{babelbib}
\usepackage{amssymb}
\usepackage{amsmath}
\usepackage{amsthm}	% hat eigene "proof"-Umgebung, macht "." nach Nummer in Theorem & Co., bietet \newtheoremstyle
\usepackage{fancyhdr}	% eigene Kopf- und Fusszeilen
\usepackage[usenames]{color}		% fuer farbige Schrift; Aufruf mit z.B.
\usepackage{enumerate}			% macht "\begin{enumerate}[(1)]"
\usepackage[latin1]{inputenc}
\usepackage[
pagebackref%, breaklinks, urlcolor=blue
]{hyperref}

\newtheoremstyle{thm}{\topsep}{\topsep}%
     {\slshape}%         Body font
     {}%         Indent amount (empty = no indent, \parindent = para indent)
     {\bfseries}% Thm head font
     {.}%        Punctuation after thm head
     { }%     Space after thm head (\newline = linebreak)
     {\thmnumber{#2.\,\,}\thmname{#1}\thmnote{\;\,(#3)}}%         Thm head spec

\newtheoremstyle{exi}{\topsep}{\topsep}%
     {}%         Body font
     {}%         Indent amount (empty = no indent, \parindent = para indent)
     {\bfseries}% Thm head font
     {.}%        Punctuation after thm head
     { }%     Space after thm head (\newline = linebreak)
     {\thmnumber{#2.\,\,}\thmname{#1}\thmnote{\;\,(#3)}}%         Thm head spec

\theoremstyle{thm}
\newtheorem{theorem}{Theorem}[section]
\newtheorem{lemma}[theorem]{Lemma}
\newtheorem{proposition}[theorem]{Proposition}
\newtheorem{corollary}[theorem]{Corollary}
\newtheorem{definition}[theorem]{Definition}

\theoremstyle{exi}
\newtheorem{example}[theorem]{Example}
\newtheorem{remark}[theorem]{Remark}

\newcommand{\ve}{\varepsilon}

\newcommand{\ee}{_\varepsilon}

\newcommand{\setN}{\mathbb{N}}

\newcommand{\setR}{\mathbb{R}}

\newcommand{\calD}{{\mathcal D}}
\newcommand{\calE}{{\mathcal E}}
\newcommand{\calG}{{\mathcal G}}
\newcommand{\calN}{{\mathcal N}}

\newcommand{\Diff}{\mathrm{D}}		% doch so definieren, damit Abstand zum naechsten Buchstaben nicht so gross wird
\newcommand{\metricd}{\mathrm{d}}
\newcommand{\mypartial}{\partial}
\DeclareMathOperator{\id}{id}

\DeclareMathOperator{\proj}{pr}

\DeclareMathOperator{\Con}{C}

\DeclareMathOperator{\supp}{supp}

\newcommand{\kugel}{\mathrm{B}}
\newcommand{\kugelz}{\kugel^\mathrm{Z}}

\newcommand{\cptin}{\subset\subset}
\newcommand{\bs}{\backslash}
\newcommand{\epsint}{{(0,1]}}

\newcommand{\deltanetz}{\delta}

\newcommand{\TT}{t}
\newcommand{\SSS}{s}

\newcommand{\tk}{T}
\newcommand{\wk}{w}
\newcommand{\sk}{S}
\newcommand{\vk}{v}
\newcommand{\rk}{T^\diamond}

\newcommand{\se}{s\ee}
\newcommand{\we}{w\ee}
\newcommand{\te}{t\ee}
\newcommand{\re}{t\ee^\diamond}
\newcommand{\veps}{v\ee}

\newcommand{\seh}{\hat s\ee}
\newcommand{\teh}{\hat t\ee}

\newcommand{\ted}{\dot t\ee}
\newcommand{\sed}{\dot s\ee}
\newcommand{\wed}{\dot w\ee}
\newcommand{\vepsd}{\dot v\ee}

\newcommand{\vepsdd}{\ddot v\ee}

\newcommand{\xki}{x^{i}}
\newcommand{\xke}{x^{1}}
\newcommand{\xkz}{x^{2}}

\newcommand{\xkid}{\dot x^{i}}

\newcommand{\xei}{x\ee^{i}}
\newcommand{\xee}{x\ee^{1}}
\newcommand{\xez}{x\ee^{2}}
\newcommand{\xej}{x\ee^{j}}

\newcommand{\xel}{x\ee^{l}}
\newcommand{\xem}{x\ee^{m}}

\newcommand{\xeid}{\dot x\ee^{i}}

\newcommand{\xeidd}{\ddot x\ee^{i}}

\newcommand{\xoi}{X^{i}}
\newcommand{\xoe}{X^{1}}
\newcommand{\xoz}{X^{2}}
\newcommand{\xok}{X^{k}}
\newcommand{\xoj}{X^{j}}

\newcommand{\vo}{V}
\newcommand{\uu}{U}

\newcommand{\xeia}[1]{\xei (\xok,#1)}

\newcommand{\xeja}[1]{\xej (\xok,#1)}

\newcommand{\xkio}{x^{i}_0}
\newcommand{\xkeo}{x^{1}_0}
\newcommand{\xkzo}{x^{2}_0}

\newcommand{\vko}{v_0}

\newcommand{\xkido}{\dot x^{i}_0}
\newcommand{\xkedo}{\dot x^{1}_0}
\newcommand{\xkzdo}{\dot x^{2}_0}
\newcommand{\vkdo}{\dot v_0}

\newcommand{\mengeQ}{Q} % vorher: W
\newcommand{\mengeP}{P} % vorher: U
\newcommand{\mengeA}{A} % vorher: A
\newcommand{\mengeG}{G} % vorher: G
\newcommand{\mengeD}{D}

\newcommand{\mengeW}{W} % vorher: Z

\newcommand{\Cinf}{{\Con^\infty}}
\newcommand{\al}{\ensuremath{\alpha}}
\newcommand{\N}{\mathbb N}
\newcommand{\pa}{\partial}

\newcommand{\mengeH}{H}
\newcommand{\hlinks}{H_\mathrm{l}}
\newcommand{\hrechts}{H_\mathrm{r}}

\newcommand{\de}{\delta}

\newcommand{\xx}{x_0,\dot x_0}

\begin{document}

\selectlanguage{british}

\title{Inversion of a ``discontinuous coordinate transformation''
 in general relativity}

\author{Evelina Erlacher\thanks{Fakult\"at f\"ur Mathematik, Universit\"at
Wien, Nordbergstr.\ 15, A-1090 Wien, Austria, \hbox{e-mail:}
evelina.erlacher@univie.ac.at,
michael.grosser@univie.ac.at}\hphantom{.},
Michael Grosser$^*$
}

\date{}

\maketitle

\begin{abstract}
\setlength{\parindent}{0pt}
In \cite{penrose2}, Penrose---in a purely formal way---introduced a
``discontinuous coordinate transformation'', which relates a continuous
representation of the metric of impulsive pp-waves to a discontinuous one. On
the basis of the invertibility concept for generalized functions developed
recently by the first author in \cite{EEinvers}, we show that this
discontinuous coordinate transformation indeed represents an invertible
generalized function in the appropriate sense.\\

{\bf Keywords:}
discontinuous coordinate transformation,
impulsive pp-wave, distributional metric, Colombeau algebra, inverse
generalized function \\

{\bf AMS Subject Classification:}
83C35, 46F30, 26B10
\end{abstract}

%%% SECTION %%%%%%%%%%%%%%%%%%%%%%%%%%%%%%%%%%%%%%%%%%%%%%%%%%%%%%%%%%%%%%%%%%%%%%%%%%%%%%%%%%%%%%%%%%%%%%%%%%%%%%%%%%%

\section{Introduction}
\label{secintro}

In general relativity, so-called impulsive pp-waves have
been described by two different metrics: In \cite[Chapter 4]{penrose2}, 
Penrose used the form
\begin{equation} \label{metric distri}
\metricd s^2 = f(x,y) \, \delta(u) \, \metricd u^2 - \metricd u \, \metricd v +
\metricd x^2 +\metricd y^2
\end{equation}
where $\de$ denotes the Dirac delta distribution. This space-time is flat
everywhere except for the null hyperplane $u=0$ where
the curvature is concentrated.
On the other hand, impulsive pp-waves have also
been described by the
continuous metric

\begin{align} \label{metric cont}
\metricd s^2 \nonumber
& = -\metricd u \, \metricd V + (1+\frac 1 2 \, \mypartial_{11} f \, u_+)^2 \,
\metricd X^2 + (1+\frac 1 2 \, \mypartial_{22} f \, u_+)^2 \, \metricd Y^2 \\
& \phantom{a} + \frac 1 2 \, \mypartial_{12} f \, \Delta f \, u_+^2 \, \metricd
X \, \metricd Y + 2 u_+ \, \mypartial_{12} f \, \metricd X \, \metricd Y + \frac
1 4 (\mypartial_{12} f)^2 \, u_+^2 \, (\metricd X^2 + \metricd Y^2)
\end{align}
(see formula (17) in \cite{ab3}) where for simplicity we have suppressed the
dependence of the
function $f$ on
its arguments, i.e., $f$ is to be read as $f(X,Y)$. $u_+$ denotes the kink
function on $\setR$, vanishing on the negative axis and acting as identity on
the positive axis.

Since both (\ref{metric distri}) and (\ref{metric cont}) aim at modelling
the same
physical (though idealized) situation, they have to be viewed
as equivalent
from a physical point of view. Therefore it seems plausible that a
coordinate transformation can be found relating $(u,x,y,v)$ and $(u,X,Y,V)$ such
that the respective substitutions transform (\ref{metric distri}) and
(\ref{metric cont}) into each other. Of course, since the coefficients in
(\ref{metric cont}) are continuous while (\ref{metric distri}) contains the
delta distribution, such a transformation cannot even be continuous; strictly
speaking, it has to change the topological structure of the manifold.

A transformation connecting (\ref{metric distri}) and
(\ref{metric cont}) indeed exists: Formally, it arises from the
``distributional geodesics'' of the metric \eqref{metric distri},
with vanishing initial speed in the $x$, $y$ and $v$-directions.  It has been
given by Penrose in
\cite{penrose2} for the special case $f(x,y)=
x^2-y^2$. For general $f$, it appears as formula (16) in \cite{ab3}
and has the form
\begin{align} \label{transf} \nonumber
u & = u, \\ 
x^i & = \xoi + \frac 1 2 \, \mypartial_i f(\xok) \, u_+, \\
v & = \vo + f(\xok) \, H(u) + \frac 1 4 \sum_{i=1}^2 \mypartial_i f (\xok)^2 \,
u_+\nonumber
\end{align}
where we write $(\xok)$ for
$(\xoe,\xoz)=(X,Y)$ and $(x^1,x^2)=(x,y)$. Due to the occurrence of the step
function (Heaviside function) $H$, this ``coordinate transformation'' obviously
is discontinuous. Formulas (\ref{metric cont}) and (\ref{transf}) have already
been alluded to implicitly in \cite[Chapter 4]{penrose2}, yet with
only the special case $f(x,y) := x^2-y^2$ actually written down. 

From a mathematical point of view, it certainly seems
desirable to embed this discontinuous transformation into a suitable theoretical
frame of generalized functions. Many important concepts which nowadays belong
to the core of rigorous standard mathematics had their origin and their
``prehistory'' in ingenious formal calculations of physicists. Distributions as
we know them today might serve as a paradigmatic example. To give a precise
meaning to the intuitive idea of a discontinuous coordinate transform, a
conceptual setting of generalized functions is required which admits
composition, hence nonlinear operations, and defining inverses. These
requirements immediately rule out linear distribution theory.

The nonlinear theory of generalized functions going back to J.F. Colombeau,
however, at least provides sufficiently broad concepts of composition
(cf.\cite[1.2.8, 1.2.29]{book}). However, there remain severe difficulties as
to developing a useful notion of inversion which mainly are due to the lack of
a sensible notion of range or image of a set under a generalized function. This
image set,
of course, would have to serve as the domain for any presumptive inverse (see
\cite[Section 1]{EEinvers} for a more extensive discussion). Even for the simple
case of a step function (occurring in (\ref{transf}), after all), the
mathematical status
of an ``inverse'' is by no means clear.

Nevertheless, Kunzinger and Steinbauer tackled this problem in
\cite{penrose} and \cite{Stdiss}; see also \cite[Section 5.4.3]{book}. They succeeded in giving a more precise
meaning to the intuitive resp.\ purely formal idea of equivalence of the two
descriptions of impulsive pp-waves by interpreting the discontinuous
transformation as the distributional shadow of a generalized transformation:
After regularizing the distributional space-time metric (\ref{metric distri}),
they applied a
generalized change of coordinates modelling the distributional one; then they
calculated the distributional shadow of the transformed generalized metric to
arrive precisely at the continuous form (\ref{metric cont}) (cf.\
\cite{penrose,book}). However,
their claim  that the
``generalized Penrose transformation''
envisaged above indeed represents a Colombeau generalized function having the
required properties as to domain and c-boundedness (see Definition
\ref{cbounded}) cannot be maintained as stated in \cite[Theorem 5.7.3 and its
proof]{book}. Furthermore, the question to what extent the generalized function
representing the discontinuous coordinate transformation is ``invertible'' could
not be answered (in fact, not even be posed in a precise sense) at that time due
to the lack of an appropriate notion of an inverse of a (Colombeau) generalized
function. Thus, Kunzinger and Steinbauer's analysis of the  generalized
coordinate transformation as an invertible Colombeau function had to remain
incomplete. At the appropriate places in the
subsequent sections we will review their achievements in more detail. 

Recently, the first author has presented a conceptual frame
for viewing certain Colombeau generalized functions as invertible, together with
necessary resp.\ sufficient conditions for invertibility (cf.\ \cite{EEinvers}).
It is the main purpose
of this paper to show that the discontinuous coordinate transformation
connecting (\ref{metric distri}) and (\ref{metric cont}) indeed represents an
invertible generalized function in the appropriate sense.

This paper is organized as follows: Section \ref{secppwaves} collects some
background information on pp-waves while Section \ref{notprelim} provides the
basic terminology for Colombeau algebras, covering, in particular, c-boundedness
and invertibility of Colombeau functions. The principal steps towards
the main result of this article are reflected by the pattern of Sections
\ref{penrose 2}--\ref{inversion}: The construction of Kunzinger and 
Steinbauer (\cite{geo2,penrose} resp.\ \cite{book}) yielding a generalized
solution
of the regularized geodesic equation
corresponding to the metric (\ref{metric distri}) 
is reviewed and complemented
in Section \ref{penrose 2}. From these geodesics, we obtain the
generalized coordinate transformation $\tk:=[(\te)\ee]$ which will be
established as a c-bounded Colombeau generalized function from
$\setR^4$ to $\setR^4$ in Section \ref{gen coord trans}. Section
\ref{injectivity} again builds on and extends results of Kunzinger and
Steinbauer  \cite{penrose, book} concerning injectivity of
$\te$ and the ``strict
non-zeroness''of the Jacobian determinant of $\te$.
Both these properties (to be satisfied on sufficiently large sets) will be
crucial
for showing the invertibility of $\tk$  in the final section.
We provide the necessary information on size and shape of the sets of
injectivity and their dependence on the relevant parameters. Now the main
difficulty in establishing the (local) invertibility of $\tk$ consists in
proving that the images of sufficiently large open sets (fixed with respect
to $\ve$) under the maps $\te$ intersect with non-empty interior, for $\ve$
small. This intersection will then serve to obtain the domains for local
inverses of the $\te$ and, in the sequel, also for the inverse generalized
function. By means of a
non-trivial result from \cite{EEinvers} on the stability of
image sets under injective continuous functions, this intersection is shown in
Section \ref{inversion} to be non-empty indeed. Putting together all the pieces,
we finally obtain local invertibility of $\tk$ in Theorem \ref{main
result}. The results of Section \ref{convergence} prepare the ground for the
final section by providing some convergence
relations needed for applying the stability theorem.

%%% SECTION %%%%%%%%%%%%%%%%%%%%%%%%%%%%%%%%%%%%%%%%%%%%%%%%%%%%%%%%%%%%%%%%%%%%%%%%%%%%%%%%%%%%%%%%%%%%%%%%%%%%%%%%%%%

\section{Plane fronted gravitational waves with
parallel rays (pp-waves)}
\label{secppwaves}

This section collects some basic facts on pp-waves, together with corresponding
references. 

The line elment of a plane fronted gravitational wave with parallel rays
(a space-time
characterized by the existence of a covariantly constant null vector field) or,
for short, a pp-wave can be written in the form 
\begin{equation} \label{brinkmann form}
\metricd s^2 = h(u,x,y) \, \metricd u^2 - \metricd u \, \metricd v + \metricd
x^2 + \metricd y^2,
\end{equation}
where $h$---the wave profile---is an arbitrary smooth function of the retarded
time coordinate $u$ and the Cartesian coordinates $x$, $y$ spanning the
wave surface.
\cite[Subsection 5.3.1]{book} surveys various aspects of pp-waves and provides
numerous references.

If the wave profile is given by $h(u,x,y)=f(x,y) \, \delta
(u)$ for $f$ an arbitrary smooth function and $\delta$ the Dirac-$\delta$ 
(cf.\ (\ref{metric distri})) the
corresponding space-times are called {\it impulsive} pp-waves. Penrose
introduced such space-times as limits of suitable
sequences of sandwich waves (cf.\ \cite{penrose0}). Moreover, they naturally
arise in a number of situations, e.g.\ as ultrarelativistic limits of boosted
black hole geometries of the Kerr-Newman familiy \cite{as,hb3,ls-kn}, as
multipole solutions of the Weyl family \cite{jiri-bsmp}, and in particle
scattering at the Planck scale \cite{v2,deharo}.

Various aspects of impulsive pp-waves have been discussed by several
authors. Let us mention \cite{ab3} and \cite{pv1} for continuous forms of
the metric, the ``scissors and paste approach'' of Penrose in \cite{penrose2}
and the work \cite{dt} of Dray and t'Hooft.

The obvious disadvantage of a description of impulsive pp-waves by
\eqref{metric distri} is the occurrence of distributional coefficients in the
metric and, consequently, also in the corresponding geodesic equations
given by
\begin{align} \label{geodesic equation} \nonumber
\ddot x^i(u) &= \frac 1 2 \, \mypartial_i f(x^1(u),x^2(u)) \, \delta (u),
\\[-10pt]
\\[-10pt]
\nonumber
\ddot v(u) &= f(x^1(u),x^2(u)) \, \dot \delta (u) + 2 \sum_{i=1}^2
\mypartial_i f(x^1(u),x^2(u)) \, \dot x^i(u) \, \delta (u),
\end{align}
(cf.\ \cite{geo} for their derivation). The right hand side of the equation for
$v$ involves
the product of $\delta$ and the Heaviside function $H$ (due to $x^i(u)$
involving the kink function $u_+$, cf.\ \cite[Theorem 5.3.3]{book}) which is not
defined in
the linear theory of distributions.
Nevertheless, attempts have been made to
solve the system  \eqref{geodesic equation} (though ill-defined) in
$\calD'$ (cf.\ \cite{fvp,herbertgeo}) by simply
setting $H \delta = \frac 1 2 \delta$. Such {\it ad hoc} multiplication rules
may work out in certain instances (such as this one, cf.\ \cite{geo,geo2,book})
but in just as many cases they will lead to considerable difficulties (cf.\
e.g.\ \cite{hajek,rn}). The nonlinear theory of generalized functions as
introduced by J.\ F.\ Colombeau (see \cite{c1,c2,MOBook,book}) provides a
setting where these problems can be overcome in a rigorous mathematical fashion
without the need for imposing such multiplication rules. Indeed, Kunzinger and
Steinbauer presented a method of treating equations such as \eqref{geodesic
equation} in a mathematically satisfactory way (see 
\cite{geo,geo2,penrose,book}): They regularized the given equations, solved them
in a
suitable Colombeau algebra and showed that the solutions indeed possess
regularization-independent distributional limits coinciding with the
distributional ``solutions'' given in \cite{fvp} and \cite{herbertgeo}.

There have been a number of further successful applications of Colombeau theory
to general relativity. For an extensive list of references, we refer to the one
at the end of the
survey article \cite{SV06} of Steinbauer and Vickers.
%containing 95 items.
The recent paper \cite{SV09} compares two different approaches to metrics of low
differentiability in general relativity.

%%% SECTION %%%%%%%%%%%%%%%%%%%%%%%%%%%%%%%%%%%%%%%%%%%%%%%%%%%%%%%%%%%%%%%%%%%%%%%%%%%%%%%%%%%%%%%%%%%%%%%%%%%%%%%%%%%

\section{Notation and preliminaries}
\label{notprelim}

For subsets $A,B$ of a topological space $X$, we write $A\subset\subset B$ if
$A$ is a compact subset of the interior $B^\circ$ of $B$. For each
non-empty open subset of $U$ of $\setR^n$ we denote by $\calD(U)$ the linear
space of test functions on $U$, i.e., of infinitely differentiable
real-valued functions having compact support in $U$. 

Concerning fundamentals of (special) Colombeau algebras, we follow
\cite[Subsection 1.2]{book}. As to inversion of generalized functions, we adopt
terminology and results from \cite{EEinvers}.

In particular, for defining the special Colombeau algebra $\calG (U)$ on
a given (non-empty) open subset $U$ of $\setR^n$, we set
$\calE (U ) :=\Con^\infty (U,\setR)^\epsint$ and

\begin{align*}
\calE_M (U ) & := \{ (u\ee)\ee \in \calE (U) \, | \, \forall \, K \cptin U \
\forall \, \alpha \in \setN_0^n \ \exists \, N \in \setN : \\
& \phantom{aaaaaaaaaaaaaaaaaaaaaaaaa} \, \sup_{x \in K} | \mypartial^\alpha u\ee
(x) | = O(\ve^{-N}) \mbox{ as } \ve \to 0 \}, \\
\calN (U ) & := \{ (u\ee)\ee \in \calE (U) \, | \, \forall \, K \cptin U \
\forall \, \alpha \in \setN_0^n \ \forall \, m \in \setN : \\
& \phantom{aaaaaaaaaaaaaaaaaaaaaaaaa} \, \sup_{x \in K} | \mypartial^\alpha u\ee
(x) | = O(\ve^m) \mbox{ as } \ve \to 0 \}.
\end{align*}
Elements of\/ $\calE_M(U)$ resp.\ $\calN(U)$ are called {\it moderate} resp.\
{\it negligible functions}. $\calE_M(U)$ is a subalgebra of $\calE (U)$, $\calN
(U)$ is an ideal in $\calE_M (U)$. The {\it special Colombeau algebra} on $U$ is
defined as
\[
\calG (U) := \calE_M (U) / \calN (U).
\]

The class of a moderate net $(u\ee)\ee$ in this quotient space will be denoted
by $[(u\ee)\ee]$.  A
generalized function on some open subset $U$ of $\setR^n$ with values in
$\setR^m$ is given as an $m$-tuple $(u_1,\cdots,u_m)\in \calG(U)^m$ of
generalized functions $u_j\in\calG(U)$ where $j=1,\cdots,m$.

The composition $v\circ u$ of two arbitrary generalized functions is not
defined, not even if $v$ is defined on the whole of $\setR^m$ (i.e., if
$u\in\calG(U)^m$ and $v\in\calG(\setR^m)^p$).
A convenient condition for $v\circ u$ to be defined is to require $u$ to be
``compactly bounded'' (c-bounded) into the domain of $v$. Since there is a
certain inconsistency in \cite{book} concerning the precise description of
c-boundedness (see \cite[Section 2]{EEinvers} for details) we
include the
explicit definition of this important property below. For a full discussion,
see again \cite[Section 2]{EEinvers}.

\begin{definition}\label{cboundednet}
Let $U$ and $V$ be open subsets of $\setR^n$ resp.\ $\setR^m$. An element
$(u\ee)\ee \in \Cinf(U,V)^{\epsint}$ is called {\it compactly
bounded (c-bounded)} if the conditions
\begin{enumerate}[(1)]
\item \label{eins}
For every $K \cptin U$ there exist $L \cptin V$ and $\ve_0 \in
\epsint$ such that $u\ee (K) \subseteq L$ for all $\ve\le\ve_0$.
\item \label{zwei}
For every $K \cptin U$ and every $\al\in\N_0^m$ there exists $N\in\N$ with
$$\sup\limits_{x\in K}|\pa^\al u\ee^j(x)|=O(\ve^{-N})$$
for all component functions $u\ee^j$ ($j=1,\dots,m$) of $u\ee$.
\end{enumerate}
are satisfied. The collection of c-bounded elements of $\Cinf(U,V)^{\epsint}$
is denoted by $\calE_M [U,V]$.
\end{definition}

Obviously, $\calE_M
[U,V]$ can be viewed as a subset of $\calE_M (U)^m$ and thus determines a
certain subset of $\calG(U)^m$.

\begin{definition}\label{cbounded}
Let $U$ and $V$ be open subsets of $\setR^n$ resp.\ $\setR^m$.
\begin{enumerate}[(1)]
\item
An element $(u\ee)\ee$ of $\calE_M (U)^m$ is called
{\it  c-bounded from $U$ into $V$} if, in fact, $(u\ee)\ee\in\calE_M[U,V]$.
\item
An element $u$ of $\calG (U)^m$ is called
{\it  c-bounded from $U$ into $V$} if it has a representative which is
c-bounded from $U$ into $V$, i.e., which is a member of $\calE_M [U,V]$.
The space of all c-bounded generalized functions from $U$ into
$V$ will be denoted by $\calG [U,V]$.
\end{enumerate}
\end{definition}

Due to the asymptotic nature of the conditions defining $\calE_M (U)$ resp.\
$\calN(U)$, the property of a generalized function $u\in\calG (U)^m$ to be
c-bounded from $U$ into $V$ is not affected if we only require the existence of
a
representative $(u\ee)\ee$ satisfying $u\ee (U)\subseteq V$ and conditions
\ref{cbounded}\,\eqref{eins} and \ref{cbounded}\,\eqref{zwei} for all $\ve$ below
some $\ve_1>0$ depending on the
net at hand.

\begin{proposition}
\label{composition}
Let $u \in \calG (U)^m$ be c-bounded into $V$ and let $v \in \calG (V)^p$, with
representatives $(u\ee)\ee$ resp.\ $(v\ee)\ee$. Then the composition
\[
v \circ u := [(v\ee \circ u\ee)\ee]
\]
is a well-defined generalized function in $\calG (U)^p$.
\end{proposition}

Next, we present the notions of invertibility introduced in \cite{EEinvers}.
\begin{definition}[Invertibility of generalized functions]
\label{inv}
Let $U$ be an open subset of $\setR^n$ and $u \in \calG (U)^n$. Let $\mengeG$ be
an open subset of $U$.
\begin{enumerate}
\item[{\bf (LI)}]
$u$ is called {\it left invertible on $\mengeG$} if
there exist $v
\in \calG (V)^n$ with $V$ an open subset of $\setR^n$ and an open set $\mengeH
\subseteq V$ such that $u|_\mengeG$ is c-bounded into $\mengeH$ and $v \circ
u|_\mengeG = \id_\mengeG$. Then $v$ is called a {\it left inverse} of $u$ on
$\mengeG$. \\[3pt]
In shorthand, $u$ is left invertible (on $\mengeG$) with left inversion data
$[\mengeG,V,v,\mengeH]$.
\item[{\bf (RI)}]
$u$ is called {\it right invertible on $\mengeG$} if there
exist $v
\in \calG (V)^n$ with $V$ an open subset of $\setR^n$ and an open set $\mengeH
\subseteq V$ such that $v|_\mengeH$ is c-bounded into $\mengeG$ and $u \circ
v|_\mengeH = \id_\mengeH$. Then $v$ is called a {\it right inverse} of $u$ on
$\mengeG$. \\[3pt]
In shorthand, $u$ is right invertible (on $\mengeG$) with right inversion data
$[\mengeG,V,$ $v,\mengeH]$.
\item[{\bf (I)}]
$u$ is called {\it invertible on $\mengeG$} if it is both left
and
right invertible on $\mengeG$ with left inversion data $[\mengeG,V,v,\hlinks ]$
and right inversion data $[\mengeG,V,v,\hrechts ]$. Then $v$ is called an {\it
inverse} of $u$ on $\mengeG$. \\[3pt]
In shorthand, $u$ is invertible (on $\mengeG$) with inversion data
$[\mengeG,V,v,\hlinks ,\hrechts ]$.
\item[{\bf (SI)}]
$u$ is called {\it strictly invertible on $\mengeG$} if it is
invertible on $\mengeG$ with inversion data $[\mengeG,V,v,\mengeH,\mengeH]$ for
some open subset $\mengeH$ of $V$. Then $v$ is called a {\it strict inverse} of
$u$ on $\mengeG$. \\[3pt]
In shorthand, $u$ is strictly invertible (on $\mengeG$) with inversion data
$[\mengeG,V,v,\mengeH]$.
\end{enumerate}
\end{definition}

Throughout this paper we will also use the phrases ``$u$ is invertible (on
$\mengeG$) by $[\mengeG,V,v,\hlinks ,\hrechts ]$'' resp.\
``$[\mengeG,V,v,\hlinks ,\hrechts ]$ is an inverse of $u$ (on $\mengeG$)''. If
we do not specify a set on which a given $u \in \calG (U)^n$ is invertible, we
always refer to invertibility on $U$, i.e.\ on the whole of its domain. The same
applies to the cases of ``left invertible'', ``right invertible'' resp.\
``strictly invertible''.

Some basic properties of the invertibility concepts introduced above are
discussed in \cite[Section 3]{EEinvers}.

\begin{definition} 
\label{d: local invertibility}
Let $U$ be an open subset of $\setR^n$ and $u \in \calG (U)^n$. We call $u$ {\it
locally (left, right) invertible} if for every point $z \in U$ there exists an
open neighbourhood $\mengeG$ of $z$ in $U$ such that $u$ is (left, right)
invertible on $\mengeG$.
\end{definition}

\begin{definition}
Let $U$ be an open subset of $\setR^n$. A moderate net $(u\ee)\ee \in \calE_M
(U)$ is called {\it strictly non-zero} if for every compact subset $K$ of $U$
there exist $C>0$, a natural number $N$ and some $\ve_0 \in \epsint$ such that
\begin{equation} \label{non-zeroness 2}
\inf_{x \in K} | u\ee (x) | \ge C \ve^N
\end{equation}
for all $\ve \le \ve_0$. An element $u$ of $\calG (U)$ is called {\it strictly
non-zero} if it possesses
a representative with this property.
\end{definition}
By \cite[Theorem 1.2.5]{book}, $u\in\calG (U)$ is strictly non-zero if and only
if there exists $v\in\calG(U)$ with $uv=1$.

%%% SECTION %%%%%%%%%%%%%%%%%%%%%%%%%%%%%%%%%%%%%%%%%%%%%%%%%%%%%%%%%%%%%%%%%%%%%%%%%%%%%%%%%%%%%%%%%%%%%%%%%%%%%%%%%%%

\section{Description of the geodesics for impulsivse pp-waves as Colombeau
generalized functions }
\label{penrose 2}

According to \cite[p.\ 1256]{penrose} resp.\ \cite[p.\ 464]{book}, the
transformation connecting the distributional and the continuous forms
of the metric (\eqref{metric distri} resp.\ \eqref{metric cont}) arises from
certain geodesics with respect to \eqref{metric distri}.
 This holds true for the regularized
resp.\ generalized versions $\te$ as well as---though only on a formal
level---for the distributional version $\TT$ given by \eqref{transf}.
In this section, therefore, we study the geodesic equations corresponding to the
regularization of the distributional metric \eqref{metric distri}, following the approach taken in \cite{geo2,penrose} resp.\ \cite{book}. We include the results obtained by
Kunzinger and Steinbauer, establishing 
existence and uniqueness of the generalized geodesics. In view of the ultimate
goal of this article, however, a more refined study of these geodesics is
required.

Following \cite{penrose}, we introduce the notion of a strict delta net as
follows:

\begin{definition} \label{strict delta net}
A {\it strict delta net} is a net $(\deltanetz\ee)\ee$ in $\calD(\setR^n)$
satisfying
\begin{enumerate}[(1)]
\item \label{bed a} \quad $\supp (\deltanetz\ee) \subseteq [-\ve,\ve]$,
\item \label{bed b} \quad $\int \deltanetz\ee(x) \, dx \to 1$ for $\ve \to 0$,
\item \label{bed c} \quad $\int |\deltanetz\ee(x)| \, dx \le C$ for some $C>0$
and small $\ve$.
\end{enumerate}
A {\it strict delta function} is a generalized function $D=[(\deltanetz\ee)\ee]
\in \calG (\setR^n)$ with $(\deltanetz\ee)\ee$ a strict delta net.
\end{definition}

Corresponding to \eqref{metric distri}, we define the generalized metric $\hat
g$ on $\setR^4$ by
\begin{equation} \label{gen metrik}
\hat \metricd s^2 = f(\xke,\xkz) \, D(u) \, \metricd u^2 -\metricd u \,
\metricd v + (\metricd \xke)^2 + (\metricd \xkz)^2,
\end{equation}
where $D$ is a strict delta function. Then, in terms of generalized functions,
the geodesic equations \eqref{geodesic equation} take the following 
form:
\begin{align} \label{geodaeten glgen 0}
\ddot x^i (u) \nonumber
& = \frac 1 2 \, \mypartial_i f(x^1(u),x^2(u)) \, D(u), \\[-10pt]
\\[-10pt]
\nonumber
\ddot v (u)
& = f(x^1(u),x^2(u)) \, \dot D (u) + 2 \sum_{i=1}^2 \mypartial_i
f(x^1(u),x^2(u)) \, \dot x^i(u) \, D (u).
\end{align}
At the level of representatives and for fixed
$\ve$, the solution of this system is obtained by means of \cite[Lemma
5.3.1]{book} (resp. \cite[Appendix]{geo}).
For the convenience of the reader and to facilitate the analysis of the
dependence of the domains of the solutions on the initial values, we state
this
lemma below. The initial conditions are chosen at $u=-1$, i.e.\ ``long before
the shock''.

\begin{lemma} \label{l: roli}
Let $g: \setR^n \to \setR^n$ and $h : \setR \to \setR^n$ be smooth and
$(\deltanetz\ee)\ee$ a net of smooth functions satisfying conditions \ref{strict
delta net}\,(\ref{bed a}) and \ref{strict delta net}\,(\ref{bed c}) as above.
For
any $\xx  \in \setR^n$ and any $\ve \in (0,1]$ consider the system
\begin{align} \label{ivp geodaeten} \nonumber
\ddot x\ee (t) & = g(x\ee(t)) \deltanetz\ee (t) + h(t) \\ 
x\ee(-1) & = x_0 \\
\nonumber
\dot x\ee (-1) & = \dot x_0.
\end{align}
Let $b>0$, $Q:= \int_{-1}^1 \int _{-1}^s |h(r)| dr \, ds$, $I:=\{ x \in \setR^n
\, | \, |x -x_0| \le b + |\dot x_0|+Q \}$ and
\[
\alpha := \min \left( \frac{b}{C \| g \|_{\infty,I} + |\dot x_0|}\,,\,\frac{1}{2
L C}\,,\,1 \right),
\]
with $L$ a Lipschitz constant for $g$ on $I$. Then (\ref{ivp geodaeten}) has a
unique smooth solution $x\ee$ on $J\ee := [-1,\alpha - \ve]$. Furthermore, for $\ve$ sufficiently small (e.g.\ $\ve \le
\frac{\alpha}{2}$) $x\ee$ is globally defined and both $(x\ee)\ee$ and $(\dot
x\ee)\ee$ are bounded on compact subsets of $\setR$, uniformly in $\ve$ for
small $\ve$.
\end{lemma}

The proof of the uniqueness part of \cite[Lemma 5.3.1]{book} actually has to be
complemented by an additional argument since, in fact, it only establishes
uniqueness of $x\ee$ as an element of $X\ee:=\{z\in\Con(J_\ve,\setR^n)\mid
|z(t)-x_0|\leq b+|\dot x_0|+Q\}$ (this is duly taken account of in
\cite[Lemma 4.2]{EEdiss}). Assuming $y\ee$ to be any solution of
\eqref{ivp geodaeten}, let $[-1,t_1]$ be the maximal
subinterval of
$[-1,\alpha-\ve]$ on which $|y\ee(t)-x_0|$ is bounded by $b+|\dot x_0|+Q$. By
integrating the differential equation twice within $[-\ve,t_1]$, the assumption
$t_1<\alpha-\ve$ leads to a contradiction. Therefore, $y\ee\in X\ee$ and,
consequently, $y\ee=x\ee$.

For fixed initial values $x_0$, $\dot x_0$ and for small $\ve$ (say,
$\ve\leq\frac{\al}{2}$, with $\al$ depending on $\dot x_0$ and $x_0$ via $I$,
$L$ and $\Vert g\Vert_{\infty,I}$), the preceding lemma ensures the existence
of a solution of the geodesic equations \eqref{geodaeten glgen 0}, defined on
$\setR$. This was exploited successfully in \cite{geo,geo2, penrose}
resp.\ \cite{book}. However, in view of our ultimate goal of establishing the
generalized coordinate transformation \eqref{transf} induced by the generalized
geodesics as an invertible generalized function in the sense of
\cite{EEinvers}, a closer analysis of the role played by $x_0$ and $\dot x_0$
in Lemma \ref{l: roli} is required: Viewing the solutions $x\ee$ of
\eqref{ivp geodaeten} as functions of $(\ve,\xx ,t)$, Lemma \ref{l:
roli} yields domains of the form
$\bigcup_{\xx \in\setR^n}(0,\frac{\al(x_0,\dot
x_0)}{2}]\times\{\xx \}\times \setR$, and bounds for $x\ee$ depending
on $\xx $ via several intermediate steps. To establish our main
result, however, we need uniformity of the domain of the solutions, as well as
uniformity of bounds for $x\ee$. Here, uniformity is to be understood as
uniformity with respect to $x_0$ and $\dot x_0$ ranging over compact subsets of
$\setR^n$. This necessary upgrading of Lemma \ref{l: roli} is accomplished by
Propositions \ref{glue} (uniformity of domains) and \ref{p: unif bded}
(uniformity of bounds) below.

The sets $I$ and $J\ee$ as well as the constants $\alpha$ and $L$ depend on the
initial values $x_0$ and $\dot x_0$. Nevertheless, they can be chosen uniformly
for $(x_0, \dot x_0)$ ranging over some compact set $K \cptin \setR^{2n}$: For
$\beta (K) := \sup_{z \in \proj_2 (K)} |z|$, set $I(K):= \proj_1 (K) +
\overline{\kugel_{b+\beta (K)+Q} (0)}$, $L(K) := \max_{z \in I(K)} \| \Diff g(z)
\|$, $\alpha (K)$ as in Lemma \ref{l: roli} (replacing $I$, $|\dot x_0|$, $L$ by
$I(K)$, $\beta (K)$, $L(K)$, respectively) and, finally, $J\ee (K) := [-1,
\alpha (K) - \ve]$.
Hence, for $\ve \le \ve (K):= \frac{\alpha (K)}{2}$ and $(\xx ) \in K$,
the solutions $x\ee(\xx )$ are globally defined. Note that 
$\beta (K)$, $I(K)$, $L(K)$ are monotonically increasing with $K$; $\al(K)$ and
$J\ee(K)$ are decreasing as $K$ increases.
By the Existence and Uniqueness Theorem for ODEs, $x\ee$ also depends smoothly
on the initial values, i.e.\ $x\ee \in \Con^\infty (K^\circ \times \setR)$ for
$K \cptin \setR^{2n}$ and $\ve \le \ve (K)$.

\begin{proposition} \label{glue}
There exists $(x\ee)\ee \in \Con^\infty (\setR^n \times \setR^n \times
\setR,\setR^n)^{(0,1]}$ such that for every $K \cptin \setR^{2n}$ there exists 
$\ve_K \in (0,1]$ such that $x\ee (\xx ,\, . \,)$ is the global solution
of \eqref{ivp geodaeten} for all $(\xx ) \in K$ and $\ve \le \ve_K$.
Moreover, $\ve_K\leq\frac{1}{2}\al(\xx )$ for all $(\xx )\in K$.
\end{proposition}

\begin{proof}
Let $(K_m)_m$ be an increasing sequence of compact subsets of
$\setR^{2n}$ satisfying $K_m \cptin K_{m+1}^\circ$ which exhausts $\setR^{2n}$.
Set $A_m := (\ve (K_{m+1}),\ve (K_m)] \times K_m$ and $A:= \bigcup_{m=1}^\infty
A_m$. Now, we may define a function $y : A \to \Con^\infty (\setR,\setR^n)$,
$(\ve,\xx ) \mapsto y\ee(\xx ,\, . \,)$  such that
$y\ee(\xx ,\, . \,)$ is the global solution of \eqref{ivp geodaeten}.
Let $\sigma_m \kern-1pt\in \calD (K_m^\circ)$ such that $0 \le \sigma_m
\le\kern-1pt 1$ and
$\sigma_m|_{K_{m-1}} \kern-2pt= 1$. For $\ve \in ( \ve(K_{m+1}) , \ve (K_m) ]$
(note
that $\ve(K_m)\searrow0$ as $m\to\infty$) we define
\[
x\ee (\xx ,t) :=
\left\{
\begin{array}{ll}
\sigma_m (\xx ) \cdot y\ee (\xx ,t), & (\xx ) \in K_m^\circ \\
0, & (\xx ) \in \setR^{2n} \setminus \supp \sigma_m
\end{array}
\right.
.
\]
Then $x\ee \in \Con^\infty (\setR^n \times \setR^n \times \setR,\setR^n)$ and
$x\ee |_{K_{m-1} \times \setR} = y\ee |_{K_{m-1} \times \setR}$. Since for $\ve
\in (0,\ve(K_m)]$ and $(\xx ) \in K_m$ the function $y\ee(x_0,\dot
x_0,\, . \,)$ is a global solution, $x\ee(\xx ,\, . \,)$ is a global
solution for $\ve \in (0,\ve(K_m)]$ and $(\xx ) \in K_{m-1}$.
Finally, for $K \cptin \setR^{2n}$ and $K \subseteq K_m$ set $\ve_K := \ve
(K_{m+1})$.
\end{proof}

We will call a net as in Proposition \ref{glue} an {\it asymptotic solution} of
the system of differential equations \eqref{ivp geodaeten}.

Next, we establish uniform boundedness of the asymptotic solution $(x\ee)\ee$ on
compact subsets of $\setR^n \times \setR^n \times \setR$ (as opposed to
uniform boundedness solely of
$t
\to x\ee(\xx ,t)$ on compact subsets of $\setR$, as yielded by Lemma
\ref{l: roli}), a crucial ingredient for our proof of moderateness of the
generalized coordinate transformation in Section \ref{penrose 3}.

\begin{proposition} \label{p: unif bded}
The asymptotic solution $(x\ee)\ee \in \Con^\infty (\setR^n \times \setR^n
\times \setR,\setR^n)^{(0,1]}$ is uniformly bounded on compact subsets of\/
$\setR^n \times \setR^n \times \setR$.
\end{proposition}

\begin{proof}
Let $K \times L \times J \cptin \setR^n \times \setR^n \times \setR$ and $\ve
\le \ve_{K \times L}$. Then
$\ve\leq\frac{1}{2}\al(\xx)$ for all $(\xx)\in K\times L$, hence
$[-1,\ve]\subseteq[-1,\al(\xx)-\ve]$. Consequently, on $K \times L \times
\setR$ the function $x\ee$ can be written as
\begin{align*}
x\ee(
& \xx ,t) = \\
& \left\{
\begin{array}{l}
\vphantom{\int_\frac{a}{b}^\frac{a}{b}} x_0 + \dot x_0 (t+1) + \int_{-1}^t
\int_{-1}^s h(r) dr \, ds, \hfill \phantom{aaa} t \in (-\infty,-1] \\
\vphantom{\int_\frac{a}{b}^\frac{a}{b}} x_0 + \dot x_0 (t+1) + \int_{-\ve}^t
\int _{-\ve}^s g(x\ee(\xx ,r)) \deltanetz\ee (r) dr \, ds \\
\vphantom{\int_\frac{a}{b}^\frac{a}{b}} \phantom{aaaa} + \int_{-1}^t \int
_{-1}^s h(r) dr \, ds, \hfill \phantom{aaa} t \in [-1,\ve] \\
\vphantom{\int_\frac{a}{b}^\frac{a}{b}} x\ee(\xx ,\ve) + \dot x\ee(x_0,\dot
x_0,\ve) (t-\ve) + \int_{\ve}^t \int _{\ve}^s h(r) dr \, ds, \hfill
\phantom{aaa} t \in [\ve,\infty)
\end{array}
\right. .
\end{align*}
Now the proof of the uniform estimates proceeds analogously to the
(straightforward) boundedness proof of \cite[Lemma 5.3.1]{book}, with
$\sup_{x_0\in K}|x_0|$, $\sup_{\dot x_0\in K}|\dot x_0|$, $I(K\times L)$
playing the respective former roles of $|x_0|$, $|\dot x_0|$, $I$.
\end{proof}

The following result of (Kunzinger and) Steinbauer establishes
existence and
uniqueness of generalized geodesics (\cite[Theorem 5.3.2]{book}; compare
\cite{geo,geo2,penrose}).

\begin{theorem}
\label{t: geodaeten}
Let $[(\deltanetz\ee)\ee]$ be a strict delta function, $f \in \Con^\infty
(\setR^2,\setR)$ and let $\xkeo,\xkedo,\xkzo,\xkzdo,\vko,\vkdo \in \setR$. Then
the system of generalized differential equations given (at the level of
representatives) by
\begin{align} \label{geodaeten glgen}
\xeidd (u) \nonumber
& = \frac 1 2 \mypartial_i f(\xee(u),\xez(u)) \, \deltanetz\ee(u)
\\[-10pt]\\[-10pt]\nonumber
\vepsdd (u)
& = f(\xee(u),\xez(u)) \, \dot \deltanetz\ee (u) + 2 \sum_{i=1}^2 \mypartial_i
f(\xee(u),\xez(u)) \, \xeid(u) \, \deltanetz\ee (u)
\end{align}
with initial conditions
\[
\xei (-1)= \xkio, \quad \xeid (-1)= \xkido, \quad \veps (-1)= \vko, \quad \vepsd
(-1) = \vkdo
\]
has a unique, c-bounded solution $\big( [(\xee)\ee],[(\xez)\ee],[(\veps)\ee]
\big) \in \calG (\setR)^3$. Hence, $\gamma : u \mapsto
([(\xee)\ee],[(\xez)\ee],[(\veps)\ee],u)(u) \in \calG [\setR,\setR^4]$ is the
unique solution to the geodesic equation for the generalized metric \eqref{gen
metrik}. Furthermore, $(\xei,\veps)$ can be chosen such as to solve
\eqref{geodaeten glgen} classically for $\ve$ sufficiently small.
\end{theorem}

The asymptotic solution constructed in Proposition \ref{glue} is a
representative of the generalized solution of \eqref{geodaeten glgen}. Observe
that the latter actually deserves the name ``solution'', despite all the
subtleties of the glueing process employed in Proposition \ref{glue}: Due to the
form of the ideal $\calN$, it is sufficient for equations to hold in $\calG$ if
they are satisfied ``only'' for small $\ve$ on compact sets on the level of
representatives.

According to \cite[5.3.3]{book} resp.\  \cite[Theorem 3]{geo2}, the
distributional limit of the solution of \eqref{geodaeten glgen}, i.e., the
distribution associated to
$\big( [(\xee)\ee],[(\xez)\ee],[(\veps)\ee]\big)$ in Theorem \ref{t:
geodaeten}
is given by
\begin{align}\label{geo assoc}
\xei(u) & \approx \xkio + \xkido \, (1+u) + \frac 1 2 \, \mypartial_i f (\xkeo +
\xkedo,\xkzo + \xkzdo) \, u_+  \nonumber\\
v\ee(u) & \approx \vko + \vkdo \, (1+u) + f (\xkeo + \xkedo,\xkzo + \xkzdo) \,
H(u) \\
& \phantom{aaaa} + \sum_{i=1}^2 \mypartial_i f (\xkeo + \xkedo,\xkzo + \xkzdo)
\, \Big( \xkido + \frac 1 4 \, \mypartial_i f (\xkeo + \xkedo,\xkzo + \xkzdo)
\Big) \, u_+.\nonumber
\end{align}
This reproduces in a rigorous way the ``solutions'' of \eqref{geodesic equation}
obtained by {\it ad hoc} multiplication rules in \cite{fvp,herbertgeo}.

%%% SECTION %%%%%%%%%%%%%%%%%%%%%%%%%%%%%%%%%%%%%%%%%%%%%%%%%%%%%%%%%%%%%%%%%%%%%%%%%%%%%%%%%%%%%%%%%%%%%%%%%%%%%%%%%%%

\section{The generalized coordinate transformation}
\label{gen coord trans}
\label{penrose 3}

In this section, we will define the generalized coordinate transformation
$T=[(\te)\ee]$ modelling \eqref{transf} and
establish its c-boundedness as an element of $\calG(\setR^4)^4$.
Following \cite[p.\ 1256]{penrose} resp.\ \cite[p.\ 464]{book}, $\te$
is obtained by taking certain geodesics of the
regularized version of \eqref{metric distri} (i.e., solutions of
\eqref{geodaeten glgen}) as new coordinate lines. More precisely, we have to
pick those geodesics having vanishing initial speed in the
$x^1$, $x^2$ and $v$-directions. Therefore we set
\begin{equation} \label{initial}
\xei (-1)= \xkio, \quad \xeid (-1)= 0, \quad \veps (-1)= \vko, \quad \vepsd (-1) = 0.
\end{equation}
Let $(\xei)\ee$ be the asymptotic solution of the first line of \eqref{geodaeten
glgen} with initial conditions \eqref{initial} obtained by Proposition
\ref{glue}. Using $\xei$ in the second line of \eqref{geodaeten glgen} yields an
asymptotic solution for the entire system of differential equations. Thus, we
may define the net of transformations $(\te)\ee$ by %\linebreak
$\te:=(u,\xee,\xez,\veps) : \setR^4 \to \setR^4$,
\[
\te:
\left(
\begin{array}{c}
\uu \\ \xok \\ \vo
\end{array}
\right)
\mapsto
\left(
\begin{array}{c}
\uu \\ \xeia{\uu} \\ \veps (\xok,\vo,\uu)
\end{array}
\right),
\]
where $(\xok) = (\xoe,\xoz)$ and $\xei$ and $\veps$ are given implicitly (with $(\xoe,\xoz)$ in a compact subset of $\setR^2$ and for sufficiently small $\ve$) by
\begin{align}\label{trafoxe}
\xei (\xok,\uu)
& = \xoi + \frac 1 2 \int_{-\ve}^\uu \int_{-\ve}^s \mypartial_i f (\xej
(\xok,r)) \, \deltanetz\ee(r) \, dr \, ds,\\[4pt]
\veps (\xok,\vo,\uu)
& = \vo\hphantom{^i} + \int_{-\ve}^\uu f(\xej (\xok,s)) \, \deltanetz\ee (s) \,
ds
\nonumber
\\
\label{trafove}
& \hphantom{abcde} + \int_{-\ve}^\uu \int_{-\ve}^s \sum_{i=1}^2 \mypartial_i f
(\xej (\xok,r)) \, \xeid (\xok,r) \, \deltanetz\ee (r) \, dr \, ds.
\end{align}

The ``discontinuous coordinate transformation'' \eqref{transf} will from now on
be denoted by  $\TT:=(u,\xke,\xkz,\vk)
: \setR^4 \to \setR^4$. Recall that it is given by
\[
\TT :
\left(\kern-1.1pt
\begin{array}{c}
\uu \\ \xok \\ \vo
\end{array}
\kern-1.1pt\right)
\mapsto
\left(
\begin{array}{rl}
u(\uu) & = \uu \\
\xki (\xok,\uu) & = \xoi + \frac 1 2 \mypartial_i f(\xok) \, \uu_+ \\
\vk (\xok,\vo,\uu) & = \vo + f(\xok) \, H(\uu) + \frac 1 4 \sum_{i=1}^2 \mypartial_i f (\xok)^2 \, \uu_+
\end{array}
\right)\kern-1pt.
\]

The following proposition provides the necessary $\calE_M$-estimates and
uniform bounds for $\te$ resp.\ its components showing, in particular, that
$(\te)\ee$ resp.\  $\tk$ are c-bounded from $\setR^4$ into $\setR^4$. Relevant
techniques of proof have essentially been developed by Kunzinger and Steinbauer:
Starting from the uniform bounds for $x\ee$ and $\dot x\ee$ provided by
Proposition \ref{p: unif bded}, derivatives
with respect to $U$ are handled by induction using the geodesic equations, while
for derivatives with respect to $X^i$ an argument involving Gronwall's Lemma is
employed (see part (iv) of the proof of the following proposition). Actually,
the latter method was used by Kunzinger and Steinbauer in a different context,
cf. \cite[p.\ 1258]{penrose} resp.\ \cite[Theorem 5.3.6]{book}; we will have to
deal with that issue below in Proposition \ref{falsch}.

Note that Theorem \ref{t: geodaeten} (due to Kunzinger and Steinbauer)
establishes (moderateness and) c-boundedness of the solutions of the geodesic
equations (hence, of the component functions of $\te$) only for fixed initual
values, i.e., by viewing the solutions as functions depending solely on the real
variable $u$. Consequently, only derivatives with respect to $U$ and only
compact sets of the form $K\times\{(\xx,v_0)\}$ (with $K\cptin \setR$) are taken
into account in the c-boundedness estimates. As mentioned already in Section
\ref{penrose 2} when discussing the application of Lemma \ref{l: roli} to the
geodesic equations, this point of view is not sufficient for our present
purpose: We definitely need to consider (the component functions of) $\te$ as
depending on four real variables simultaneously.

In what follows, we will use the following abbreviations for partial
differentiation operators:
$\pa_U:=\frac{\pa}{\pa U}$; $\pa_{\xoj}:=\frac{\pa}{\pa X^j}$;
$\pa_X^\al:=\pa_{\xoe}^{\al_1}\pa_{\xoz}^{\al_2}$ where $j\in\{1,2\}$ and
$\al=(\al_1,\al_2)\in\setN^2_0$.
For a detailed proof of the following proposition we refer to \cite[Proposition
4.7]{EEdiss}.

\begin{proposition} \label{p: t mod}
$\tk = [(\te)\ee]$ is an element of $\calG [\setR^4,\setR^4]$.
Furthermore,
$( \pa_X^\al  \xei )\ee$
and
$( \pa_X^\al  \pa_U\xei )\ee$
are c-bounded from
$\setR^3$ into $\setR$ for $\al\in \setN_0^2$ and $i=1,2$.
\end{proposition}

\begin{proof}
The proof is split into parts (i)-(viii) which altogether establish all claims.
All estimates are to be understood to hold true for small $\ve$.

(i) By Proposition \ref{p: unif bded}, $\xei$ is uniformly bounded
on compact subsets of $\setR^4$.

(ii) Differentiating \eqref{trafoxe} with respect to $U$ leads to a
uniform estimate on compact subsets of $\setR^4$ also for $\pa_U\xei$.

(iii) Now the geodesic equation for $\xei$ inductively yields
$\calE_M$-estimates for $\pa_U^k\xei$ for $k\geq2$.

(iv) In order to estimate
\begin{equation} \label{bloed}
\mypartial_{\xoj} \xei (\xok,\uu)
= \delta^i_j + \frac 1 2 \int_{-\ve}^\uu \int_{-\ve}^s \sum_{m=1}^2 \mypartial_m
\mypartial_i f (\xel (\xok,r)) \, \mypartial_{\xoj} \xem (\xok,r) \,
\deltanetz\ee(r) \, dr ds 
\end{equation}
on some compact subset $K \times [-1,u_0]$ of $\setR^3$  we define (following
\cite{penrose})
$$g\ee(K,u_0) := \sup
\big\{
\sum_{i=1}^2 \big| \mypartial_{\xoj} \xei (\xok,\uu) \big|
\, | \,
(\xok,\uu) \in K \times [-1,u_0], \, j =1,2
\big\}.$$
From \eqref{bloed} we obtain an estimate of the form
\begin{align*}
|g\ee(K,u_0)|
& \le 1 + C \, C_{K,u_0} \int_{-\ve}^{u_0} |g\ee(K,s)| \, ds,
\end{align*}
where $C_{K,u_0}$ is the supremum of $| \mypartial_i \mypartial_j
f(\xel(\xok,\uu)) |$ with $(\xok,\uu)$ ranging over $K \times [-1,u_0]$,
$i,j \in \{ 1,2 \}$ and $C$ is the constant from \ref{strict delta
net}\,\eqref{bed c}.
Gronwall's Lemma now implies that for small $\ve$, 
$\mypartial_{\xoj} \xei$ remains uniformly bounded on compact subsets of
$\setR^3$ (note that $\mypartial_{\xoj} \xei
(\xok,\uu) = \delta^i_j$ for $\uu \le -\ve$).

(v) By induction, we obtain uniform estimates for higher order derivatives with
respect to $X$:
For $\alpha \in \setN_0^2$ with $|\alpha|\ge 2$, a somewhat involved calculation
gives
\[
|\mypartial_{\xoj} \mypartial_X^\alpha \xeia{\uu}|
\le C_1 + \frac 1 2 C_2 \int_{-\ve}^\uu \int_{-\ve}^s |\deltanetz\ee(r)|
\sum_{m=1}^2 |\mypartial_{\xoj} \mypartial_X^\alpha \xem (\xok,r)| \, dr \, ds
\]
where $(\xok,\uu)$ ranges over some compact set, $C_1,C_2$ are positive
constants and $\ve$ is sufficiently small. Estimating in a way similar to the
case $|\alpha|=1$ yields that
also $ \mypartial_{\xoj} \mypartial_X^\alpha \xei $ is
uniformly bounded on compact subsets of $\setR^3$.

(vi) From
\begin{equation} \label{fast fertig 2}
\mypartial_X^\alpha \mypartial_\uu \xeia{\uu}
= \frac 1 2 \int_{-\ve}^\uu \mypartial_X^\alpha \big( \mypartial_i
f(\xeja{s}) \big) \deltanetz\ee(s) \, ds
\end{equation}
one obtains uniform bounds (on compact sets) for $\pa^\al_X\pa_U\xei$.

(vii) $\calE_M$-estimates for $\pa^\al_X\pa_U^m\xei$ (where $m\geq 2$) follow
inductively by differentiating \eqref{fast fertig 2} with respect to $U$.

(viii) C-boundedness of $(\veps)\ee$ is a direct consequence
of the c-boundedness of $(\xei)\ee$ and
$(\mypartial_\uu \xei)\ee$, taking into account condition \ref{strict delta
net}\,\eqref{bed c} on $\deltanetz\ee$.

Altogether, (i)--(vii) result in $(\xei)\ee$ being c-bounded from $\setR^4$ to
$\setR$ (including $V$ as a dummy variable); (iv)--(vii) establish 
$(\pa^\al_X\xei)\ee$ as c-bounded, as (vi)--(vii) do for
$(\pa^\al_X\pa_U\xei)\ee$. C-boundedness of $(\veps)\ee$, finally, is
accomplished by (viii).
\end{proof}

The last part of \cite[Theorem 5.3.6]{book} seems to partly anticipate
Proposition \ref{p: t mod} by stating
that $T=[(\te)\ee]$ is c-bounded from some open subset of $\setR^4$ into
$\setR^4$. In the respective proof, however, this claim is covered solely by the
remark ``[\dots] is immediate from Lemma 5.3.1''(Lemma
\ref{l: roli} in this article). In view of the length
of the (already fairly compact) proof of Proposition \ref{p: t mod}, the remark
from the
proof of \cite[Theorem 5.3.6]{book} cited above suggests an oversight on the
part of the
authors. Be that as it may, we decided to include a condensed
version of
the proof of Proposition \ref{p: t mod} for the sake of completeness.

%%% SECTION %%%%%%%%%%%%%%%%%%%%%%%%%%%%%%%%%%%%%%%%%%%%%%%%%%%%%%%%%%%%%%%%%%%%%%%%%%%%%%%%%%%%%%%%%%%%%%%%%%%%%%%%%%%

\section{Injectivity}
\label{injectivity}
In this section we will show injectivity of the ``discontinuous coordinate
transformation'' $\TT$ and the functions $\te$ of the generalized
transformation, each on suitable subsets of $\setR^4$. Moreover, the Jacobian
determinant of $\te$ will be proved to be strictly non-zero on these sets.

For classical functions, injectivity obviously is a necessary condition for
being invertible. In the appropriate sense, this also holds true for
(Colombeau) generalized functions (\cite[Proposition 4.5]{EEinvers}). Hence it
is natural to have the present section in this article. Yet there is another
reason, much deeper than the previous one, why injectivity is needed to
establish $\tk$ as invertible:
In order to prove ``asymptotic stability'' of image sets
under $(\te)\ee$, i.e., to show
that there exist open sets $\mengeP$ such that the family $(\te(\mengeP))\ee$
intersects with non-empty interior, we are going to employ a stability
theorem due to the first author (\cite[Theorem 4.6]{EEinvers}) which, in
turn, is based on a theorem of
Brouwer (\cite[Theorem 7.12]{mad}). Brouwer's theorem has
injectivity of the functions involved among its assumptions.

In order to turn four-vectors into three-vectors we introduce the following
notation: For any $x=(x^1,\ldots,x^n)\in\setR^n$ ($n \ge 2$), set $\hat
x:=(x^1,\ldots,x^{n-1})$, and for functions $f$ from some set into $\setR^n$,
$f=(f^1,\ldots,f^n)$, set $\hat f:=(f^1,\ldots,f^{n-1})$. If
$f=(f^1,\ldots,f^n)$ is a function of $x=(x^1,\ldots,x^n)$ with only $f^n$
actually depending on $x^n$, we will not formally distinguish between $\hat f$
considered as a function of $x$ ($n$ variables) and of $\hat x$ ($n-1$
variables). The respective meaning will be clear from the context.

For $\hat\TT$ (hence for $\TT$), injectivity on some open set containing the
half space $(-\infty,0] \times \setR^2$ is established by the following lemma,
setting $g=\frac 1 2 \Diff f$. Two examples will then show that in the special
case $f(X,Y)=X^2-Y^2$ considered by Penrose in \cite{penrose2} such a
neighbourhood is given by $(-\infty,1) \times \setR^2$, whereas for general
(smooth) $f$ a rectangular set of injectivity, i.e.\ one of the form
$(-\alpha,\beta) \times \setR^2$ with $\al,\beta>0$, does not necessarily
exist.

\begin{lemma} \label{schlange}
Let 
\[
\begin{array}{crcl}
F: & (-a,b) \times \setR^n & \to & (-a,b) \times \setR^n \\
& & & \\
& \left(
\begin{array}{c}
\uu \\ X
\end{array}
\right)
& \mapsto
& \left(
\begin{array}{l}
\uu \\
X + g(X) \, \uu_+
\end{array}
\right).
\end{array}
\]
where $a,b \in \setR^+ \cup \{ \infty \}$ and $g \in \Con^1 (\setR^n, \setR^n)$. Then there exists an open set $\mengeW$ containing $(-a,0] \times \setR^n$ such that $F|_\mengeW$ is injective.
\end{lemma}

\begin{proof}
For $X \in \setR^n$ define $h(X):=\sup_{z \in \overline{\kugel_{|X|}(0)}} \|
\Diff g(z) \|$. The function $h$ is continuous, non-negative and non-decreasing
with $|X|$. Now set
\[
\mengeW:= \left\{ (\uu,X) \in (-a,b) \times \setR^n \,
\Big| \, -a < \uu < \min \Big( b, \frac{1}{h(X)} \Big) \right\}
\]
(here we use
the convention $\frac 1 0 := \infty$). Let $(\uu_1,X_1)$, $(\uu_2,X_2) \in
\mengeW$ and \linebreak 
$F (\uu_1,X_1)=F (\uu_2,X_2)$. Then $\uu_1=\uu_2=:\uu$
and $\uu <
\frac{1}{h(X_i)}$ for $i=1,2$. For $\uu \le 0$, we immediately obtain $X_1=X_2$.
Now let $\uu>0$ and assume $X_1 \not= X_2$ with $|X_1| \ge |X_2|$, w.l.o.g.
Then
\[
|X_1-X_2|
= \uu \cdot | g(X_2)-g(X_1) |
\le \uu \cdot \sup_{z \in \overline{\kugel_{|X_1|}(0)}} \| \Diff g(z) \| \cdot |X_2-X_1|
< |X_2-X_1|,
\]
thereby concluding the proof by contradiction.
\end{proof}

In the following two examples, we consider $F$ as in Lemma \ref{schlange}, with
$g$ being given as $\frac 1 2 \, \Diff f$ for certain functions $f: \setR^2 \to
\setR$. The map $F$ then represents $\hat \TT$, i.e. the first three
components of $\TT$ corresponding to the function $f$ at hand.

\begin{example} \label{bsp1}
Let $f: \setR^2 \to \setR$, $f(X,Y) := X^2-Y^2$. This special case was
considered by Penrose in \cite{penrose2} (cp.\ also \cite{book}, components
$1$,$2$,$4$ of ($5.45$) on p.\ 463). In this case, an easy computation shows
that $\hat \TT$ is injective (even) on $(-\infty,1) \times \setR^2$. The value
$1$
is maximal since $\hat \TT (1,X,Y_1) =(1,2X,0)= \hat \TT (1,X,Y_2)$ for all
$X,Y_1,Y_2 \in \setR$.
\end{example}

\begin{example} \label{bsp2}
Let $f: \setR^2 \to \setR$, $f(X,Y) := -\frac 1 2 (X^4+Y^4)$. For every $\eta>0$ the function $\hat \TT$ is non-injective on $\{ \eta \} \times \setR^2$ since $(\eta,0,0)=\hat \TT (\eta,0,0) = \hat \TT (\eta, \frac{1}{\sqrt{\eta}}, \frac{1}{\sqrt{\eta}}) = \hat \TT (\eta, -\frac{1}{\sqrt{\eta}}, -\frac{1}{\sqrt{\eta}})$. Hence, on every set of the form $(-\alpha,\beta) \times \setR^2$ ($\alpha,\beta>0$), $\hat \TT$ is non-injective. However, $\hat \TT$ is injective on $\mengeW= \{ (\uu,X,Y) \, | \, \uu < \frac{1}{3}(X^2+Y^2)^{-1} \}$.
\end{example}

We now turn to the question of injectivity of the generalized coordinate
transformation. In \cite{penrose} (cf.\ also \cite[Theorem 5.3.6]{book})
Kunzinger and Steinbauer claim that for sufficiently small $\ve$, the
functions $\te$ are diffeomorphisms on a suitable {\em rectangular} open
subset $\Omega$ of $\setR^4$ containing the shock hyperplane $\uu =0$. In
their proof, they
employ a global univalence theorem of Gale and
Nikaido, stating that any differentiable function $F: \Omega \to
\setR^n$,
where $\Omega$ is a closed rectangular region in $\setR^n$, is injective if all
principal minors of its Jacobian $J(x)$ are positive (see \cite{uni}). However,
a closer look at the proofs in  \cite{penrose} resp.\ of \cite[Theorem
5.3.6]{book} reveals that the condition of Gale and
Nikaido's Theorem 
in fact is established only on sets of the form $(-\infty,\eta \, ]
\times K
\times \setR$ for sufficiently small $\ve$, say $\ve \le \ve_0$, where $K$ is a
compact subset of $\setR^2$ and $\eta$ and $\ve_0$ both depend on $K$.
Furthermore, they use uniform boundedness of $(\xei)\ee$ on compact subsets
of $\setR^4$ (Proposition \ref{p: unif bded}) whereas \cite[Lemma 5.3.1]{book}
(Lemma \ref{l: roli} above) only provides boundedness on compact
subsets of $\setR$ for fixed initial values $\xkio$ and $\xkido$.

Therefore, we restate Theorem 5.3.6 of \cite{book}, keeping only those claims
which are actually shown in \cite{penrose} resp.\ \cite{book} and
complementing it with a sketch of proof.

\begin{proposition} \label{falsch}
For every $K \cptin \setR^2$ and $\delta>0$ there exist $\eta>0$ and $\ve_0 \in (0,1]$ such that every principal minor of $\Diff \te(\uu,\xoi,\vo)$ stays in $(1-\delta,1+\delta)$ for all $(\uu,\xoi,\vo) \in (-\infty,\eta \, ] \times K \times \setR$ and $\ve \le \ve_0$. In particular, $\det \circ \, \Diff \tk$ is strictly non-zero on $(-\infty,\eta \, ] \times K \times \setR$ and every principal minor of $\Diff \te(\uu,\xoi,\vo)$ is positive for $(\uu,\xoi,\vo) \in (-\infty,\eta \, ] \times K \times \setR$ and $\ve \le \ve_0$.
\end{proposition}

\begin{proof}
We have to find estimates for $\frac{\mypartial \xei}{\mypartial \xoj}
(\xok,\uu)-\delta^i_j$; the partial derivative can be written as in
\eqref{bloed}.
Noting that $\frac{\mypartial \xei}{\mypartial \xoj} (\xok,\uu) = \delta^i_j$
for $\uu \le -\ve$, we obtain
\begin{align} \label{hier}
\bigg| \frac{\mypartial \xei}{\mypartial \xoj} (\xok,\uu) - \delta^i_j \bigg|
& \le C \, C_{K,1} \, C_1 \, (\uu+\ve)_+
\end{align}
for $(\xok,\uu)  \in K \times (-\infty,1]$ and sufficiently small $\ve$. Here,
 $C_1$ is a constant chosen according to the c-boundedness of $\big(
\frac{\mypartial \xei}{\mypartial \xoj} \big)\ee$ and $C_{K,1}$ has the same
meaning as in the proof of Proposition \ref{p: t mod}. Thus the supremum of the
left hand side of \eqref{hier} for $(\xok,\uu) \in K \times (-\infty,\eta]$
stays arbitrarily close to $0$ for all $\ve \le \ve_0$ if $\eta>0$ and $\ve_0
\in (0,1]$ are chosen accordingly.
\end{proof}

We will say a smooth net $(u\ee)\ee : (-a,b) \times \setR^n \times \setR \to
(-a,b) \times \setR^n \times \setR$ (for $a,b \in \setR^+ \cup \{ \infty \}$)
has {\it property (E)}\/ if for every compact subset $K$ of $\setR^n$ there
exist $\alpha \in (0,b)$ and $\ve_0 \in (0,1]$ such that $u\ee$ is injective on
$(-a,\alpha] \times K \times \setR$ for all $\ve \le \ve_0$. The net $(u\ee)\ee$
is said to have {\it property (E+)}\/ if for every compact subset $K$ of
$\setR^n$ there exist $\alpha \in (0,b)$ and $\ve_0 \in (0,1]$ such that $u\ee$
is injective on $(-a,\alpha] \times K \times \setR$ and $(\det \circ \Diff
u\ee)\ee$ is strictly non-zero, uniformly on $(-a,\alpha] \times K \times \setR$
for all $\ve \le \ve_0$, i.e. an estimate as \eqref{non-zeroness 2}
holds for all $(\uu,X,V) \in
(-a,\alpha] \times K \times \setR$.

Combining the preceding proposition and the univalence theorem of Gale and
Nikaido, it follows that $(\te)\ee$ has property (E+). By \cite[Theorem
3.59]{EEdiss}, this is already sufficient for $\tk$ to be left invertible in
the sense of Definition \ref{inv}:
\begin{corollary}
For every open relatively compact subset $W$ of $\setR^2$ there exists some
$\alpha>0$ such that for all $\beta>0$ and for all bounded open intervals $I$
the generalized function $\tk$ is left invertible on $(-\beta,\alpha) \times W
\times I$.
\end{corollary}

%%% SECTION %%%%%%%%%%%%%%%%%%%%%%%%%%%%%%%%%%%%%%%%%%%%%%%%%%%%%%%%%%%%%%%%%%%%%%%%%%%%%%%%%%%%%%%%%%%%%%%%%%%%%%%%%%%

\section{Uniform convergence}
\label{convergence}
The key idea for showing that the images of certain sets
under the $\te$ intersect with non-empty interior consists in observing that if
$\te$ stays close enough to $\TT$, then also the image of some set $W$ under
$\te$ stays close to $\TT (W)$. Therefore, convergence of $(\te)\ee$ to $\TT$
as $\ve \to 0$ in some sense might be useful. The last statement of
\cite[Theorem 5.3.3]{book} (cf.\ also \cite[Theorem 3]{geo2}) tells us that
$\xei(\, . \, ,\xoe,\xoz,\vo)$ converges to $\xki(\, . \, ,\xoe,\xoz,\vo)$
as $\ve \to 0$, yet only in the sense of uniform convergence on compact subsets
of $\setR$ for fixed $(\xoe,\xoz,\vo) \in \setR^3$. In contrast, we will need
(and establish in the sequel) uniform convergence of $(\xei)\ee$ to $\xki$ on
arbitrary compact subsets of $\setR^4$. Obviously, this is impossible for
$\veps$ since $\vk$ is discontinuous. However, cutting out the part of $\veps$
converging (pointwise for $\uu \not= 0$) to the term involving the Heaviside
function, we again can prove uniform convergence on arbitrary compact sets. To
this end, we define
\begin{align*}
\wk (\xok,\vo,\uu) := &\vo + \frac 1 4 \sum_{i=1}^2 \mypartial_i f (\xok)^2 \,
\uu_+,
\\
\we (\xok,\vo,\uu) := &\vo + \int_{-\ve}^\uu \int_{-\ve}^s \sum_{i=1}^2
\mypartial_i f (\xej(\xok,r)) \, \xeid (\xok,r) \, \deltanetz\ee (r) \, dr \,
ds.
\end{align*}
Furthermore, let $\SSS := (u,\xke,\xkz,\wk)$ and $\se :=
(u,\xee,\xez,\we)$.
Obviously, $\hat \TT = \hat \SSS$, implying that also $\hat \SSS$ is injective
on some open set containing the half space $(-\infty,0] \times \setR^2$.
Moreover, since all principal minors of $\Diff \te$ are independent of the
derivatives of $\veps$, Proposition \ref{falsch} also holds for $(\se)\ee$.
Therefore, also $(\se)\ee$ has property (E+).

In a first step, we state that $\ted \to \dot \TT$
and, due to the same proof, $\sed \to \dot \SSS$, uniformly on compact subsets
of $(\setR \bs \{ 0 \}) \times \setR^3$ for $\ve \to 0$. 
The proof proceeds along the same lines as the proof of \cite[Theorem
5.3.3]{book}. For the detailed argument we refer to
\cite[Lemma 4.13]{EEdiss}.

The formal similarity of the respective proofs of
\cite[Theorem 5.3.3]{book} and Proposition \ref{konv abl} is owed to the fact
that the former determines the limits of the right hand sides of
\eqref{geodaeten glgen}, integrated against a test function $\psi$, while the
latter establishes the limits (uniformly on compact sets) of the derivatives of
the right hand sides of \eqref{trafoxe} and \eqref{trafove}. Now, in fact,
\eqref{geodaeten
glgen} is but the second derivative of \eqref{trafoxe}\eqref{trafove}.

\begin{proposition} \label{konv abl}
$\ted \to \dot \TT$ as $\ve \to 0$, uniformly on compact subsets of $(\setR \bs \{ 0 \}) \times \setR^3$.
\end{proposition}

In order to pass from $\sed \to \dot \SSS$ to $\se \to  \SSS$, we employ the
following auxiliary result \cite[Lemma 4.14]{EEdiss}.

\begin{lemma} \label{l: glm konv}
Let $f\ee,f \in \Con (\setR^n,\setR)$ (for $\ve \in (0,1]$). Suppose that $\mypartial_n f\ee (x,t)$ and $\mypartial_n f (x,t)$ exist for all $(x,t) \in \setR^{n-1} \times (\setR \bs \{ 0 \})$ and that $\mypartial_n f\ee (x,\, . \,)$ and $\mypartial_n f (x,\, . \,)$ are piecewise continuous (with one-sided limits existing) for all $x \in \setR^{n-1}$. Let $c \in \setR$ with $c<0$. If
\begin{enumerate}[(1)]
\item \label{konv oben}
$f\ee \to f$ for $\ve \to 0$ uniformly on $K \times \{ c \}$ for all $K \cptin \setR^{n-1}$,
\item \label{konv weg von null}
$\mypartial_n f\ee \to \mypartial_n f$ for $\ve \to 0$ uniformly on compact subsets of $\setR^{n-1} \times (\setR \bs \{ 0 \})$, and
\item \label{abl bounded}
$\| \mypartial_n f\ee - \mypartial_n f \|_{\infty,K \times ([-d,d] \bs \{ 0 \})}$ is uniformly bounded for any compact set $K \cptin \setR^{n-1}$ and some $d>0$,
\end{enumerate}
then $f\ee \to f$ for $\ve \to 0$ uniformly on arbitrary compact subsets of $\setR^n$.
\end{lemma}

Now we are ready to prove

\begin{proposition}\label{konv fkt}
$\se \to \SSS$ for $\ve \to 0$ uniformly on compact subsets of $\setR^4$.
\end{proposition}

\begin{proof}
We show that for each component function of $\se$ the conditions of Lemma \ref{l: glm konv} are satisfied with respect to $\SSS$. The symbol $\mypartial_n$ in Lemma \ref{l: glm konv}, if applied to $\xei$ resp.\ $\we$, is understood to denote the derivatives of $\xei$ resp.\ $\we$ with respect to $\uu$.

$\xeid$ resp.\ $\wed$ are smooth on $\setR^3$ resp.\ $\setR^4$, $\xki$ and $\wk$
are smooth on $\setR^2 \times (\setR \bs \{ 0 \})$ resp.\ $\setR^3 \times (\setR
\bs \{ 0 \})$. $\xkid (\xoe,\xoz,\, . \,)$ and $\dot \wk(\xoe,\xoz,\vo,\, . \,)$
are piecewise continuous for all $(\xoe,\xoz) \in \setR^2$ resp.\
$(\xoe,\xoz,\vo) \in \setR^3$. For $\uu =-1$ the integral terms of $\xei (\, .
\, ,\, . \, , \uu)$ and $\we (\, . \, ,\, . \, ,\, . \, , \uu)$ vanish and
$\xei=\xki$ and $\we=\wk$. Hence, condition \eqref{konv oben} is satisfied. By
Proposition \ref{konv abl}, $\xeid \to \xkid$ and $\wed \to \dot \wk$ for $\ve
\to 0$ uniformly on compact subsets of $\setR^2 \times (\setR \bs \{ 0 \})$
resp.\ $\setR^3 \times (\setR \bs \{ 0 \})$, i.e.\ they satisfy condition
\eqref{konv weg von null}. Finally, by Theorem \ref{t: geodaeten}, $\xeid$ is
uniformly bounded on compact sets and, therefore, this is also true for $\wed$.
Since both $\xkid$ and $\dot \wk$ are bounded on any bounded subset of $\setR^2
\times (\setR \bs \{ 0 \})$ resp.\ $\setR^3 \times (\setR \bs \{ 0 \})$, also
condition \eqref{abl bounded} is satisfied and the claim follows.
\end{proof}

%%% SECTION %%%%%%%%%%%%%%%%%%%%%%%%%%%%%%%%%%%%%%%%%%%%%%%%%%%%%%%%%%%%%%%%%%%%%%%%%%%%%%%%%%%%%%%%%%%%%%%%%%%%%%%%%%%

\section{Inversion of the generalized coordinate transformation}
\label{inversion}

Finally, we turn to establishing local invertibility of the generalized
coordinate transformation $\tk$. The core of the proof consists in
showing that there exist open sets $\mengeP$ such that, for $\ve$ small, the
intersection of the $\te(\mengeP)$ has non-empty interior. The sets $\mengeP$
can be chosen such as to contain arbitrarily large (bounded) portions of the
left half space $\uu \le 0$.

The achievements of Kunzinger and Steinbauer in the context of inverting $T$
have already been discussed in some detail in previous sections; recall, in
particular, what has been said in Sections \ref{secintro}, \ref{penrose 2} and
\ref{injectivity}.

In the sequel, we will often have to
make use of cylinders rather than balls. Therefore, for $x=(\hat x,x^n) \in
\setR^n$, let $\kugelz_{\delta,\eta}(x)$ denote the cylinder $\kugel_\delta
(\hat x) \times (x^n-\eta,x^n+\eta)$.
Theorem \ref{t:  penrose-thm allg} below, being one of this section's
prominent technical tools, arises as a slightly modified version of
{\cite[Theorem 4.6]{EEinvers} where the open balls $\kugel_\delta(0)$ are
replaced by cylinders $\kugelz_{\delta,\eta}(0)$. We leave it to the reader to
adapt the proof of
\cite[Theorem 4.6]{EEinvers} to the case
of cylinders. Roughly speaking, this ``stability theorem'' establishes a
kind of continuous dependence of connected parts $f(A)$ of the image set $f(U)$
on the function $f$.} 

\begin{theorem} \label{t: penrose-thm allg}
Let $U$ be an open subset of $\setR^n$, $f,g \in \Con (U, \setR^n)$ both
injective and $W$ a connected open subset of $\setR^n$ with $\overline{W} \cptin
f(U)$. Choose $y \in W$ and $\delta,\eta >0$ with $y + \kugelz_{\delta,\eta} (0)
\subseteq W$ such that the closure of $W_{\delta,\eta} := W +
\kugelz_{\delta,\eta} (0)$ is still a subset of $f(U)$. 
If, for $A:=f^{-1} (\overline{W_{\delta,\eta}})$ and $f = (\hat f,f^{n})$ resp.\ $g = (\hat g,g^{n})$, both
\[
\| \hat g - \hat f \|_{\infty,A} < \delta \quad \mbox{and} \quad \| g^{n} - f^{n} \|_{\infty,A} < \eta
\]
hold, then
\[
\overline{W} \subseteq g(A)^\circ.
\]
\end{theorem}

Now we are ready to prove that the domains of suitable inverses of the $\te$
intersect with non-empty interior. The following theorem yields the desired
result for an entire class of c-bounded nets (also denoted by $(\te)\ee$) of
smooth functions of which our particular $(\te)\ee$ at hand is but a special
case.

\begin{theorem} \label{ich bin die groesste!}
Let $a,b \in \setR^+ \cup \{ \infty \}$. Let the functions $\te$, $\se$ (for every $\ve \in (0,1]$) and $\SSS$ satisfy the following assumptions: \vspace*{-.5mm}
\begin{enumerate}[(1)]
\item \label{penrose eins}
$
\begin{array}[t]{crcl}
%\phantom{\SSS \se} 
\!\! \te: & \!\!\! (-a,b) \times \setR^n \times \setR & \!\!\! \to & \!\!\!
(-a,b) \times \setR^n \times \setR \\
 & & & \\
 & \left(
\begin{array}{c}
\uu \\ X \\ V
\end{array}
\right)
& \!\!\! \mapsto
& \!\!\! \left(
\begin{array}{lcl}
u(\uu) & \!\!\! := & \!\!\! \uu \\
x\ee(\uu,X) & & \\
\veps(\uu,X,V) & \!\!\! := & \!\!\! V+g\ee(\uu,X)+h\ee(\uu,X)
\end{array}
\! \right)
\end{array}
$ \\[.7\baselineskip]
where $x\ee \in \Con^\infty ((-a,b) \times \setR^n,\setR^n)$ and $g\ee,h\ee \in
\Con^\infty ((-a,b) \times \setR^n,\setR)$. Assume that $(\te)\ee$ has property
(E), i.e.\ that for every compact subset $K$ of $\setR^{n}$ there exist $\alpha
\in (0,b)$ and $\ve' \in (0,1]$ such that $\te$ is injective on $(-a,\alpha]
\times K \times \setR$ for all $\ve \le \ve'$. Furthermore, suppose that
$(h\ee)\ee$ is uniformly bounded on compact subsets of $(-a,b) \times \setR^n$.

\item
$
\begin{array}[t]{crcl}
\!\! \se: & \!\!\! (-a,b) \times \setR^n \times \setR & \!\!\! \to & \!\!\!
(-a,b) \times \setR^n \times \setR \\
& & & \\
&\left(
\begin{array}{c}
\uu \\ X \\ V
\end{array}
\right)
& \!\!\! \mapsto
& \!\!\! \left(
\begin{array}{lcl}
u(\uu) & \!\!\!\! \,= & \!\!\! \uu \\
x\ee(\uu,X) & & \\
\we(\uu,X,V) & \!\!\!\! := & \!\!\! V+g\ee(\uu,X)
\end{array}
\right) .
\end{array}
$ \\[.7\baselineskip]
By \eqref{penrose eins}, $\se$ is smooth. Suppose that also $(\se)\ee$ has
property (E).

\item
$
\begin{array}[t]{crcl}
\!\! \SSS \,\, : & \!\!\! (-a,b) \times \setR^n \times \setR & \!\!\! \to &
\!\!\! (-a,b) \times \setR^n \times \setR \\
& & & \\
& \left(
\begin{array}{c}
\uu \\ X \\ V
\end{array}
\right)
& \!\!\! \mapsto
& \!\!\! \left(
\begin{array}{lcl}
u(\uu) & \!\! \,= & \!\!\! \uu \\
x(\uu,X) & & \\
\wk(\uu,X,V) & \!\! := & \!\!\! V+g(\uu,X)
\end{array}
\right)
\end{array}
$ \\[.7\baselineskip]
where $x \in \Con ((-a,b) \times \setR^n, \setR^n)$ and $g \in \Con ((-a,b)
\times \setR^n, \setR)$. Assume that for $\hat \SSS:=(u,x) : (-a,b) \times
\setR^n \to (-a,b) \times \setR^n$, there exists some open set $\mengeW$
containing $(-a,0] \times \setR^n$ such that $\hat \SSS|_\mengeW$ is injective.
\end{enumerate}
\vspace*{-.4\baselineskip}
Finally, suppose $\se \to \SSS$ for $\ve \to 0$ uniformly on compact sets.
\vspace*{.3\baselineskip}

Then the following holds: For every $p$ on the hyperplane $\uu =0$ there exist open neighbourhoods $\mengeP$ of $p$ with $\mengeP \subseteq \mengeW \times \setR$ and $\mengeQ$ of $q:= \SSS (p)$ with $\mengeQ \subseteq \SSS(\mengeW \times \setR)$, and some $\ve_0 \in (0,1]$ such that
\begin{equation*}
\overline{\mengeQ} \subseteq \te (\mengeP)
\end{equation*}
for all $\ve \le \ve_0$.
\end{theorem}

\begin{proof}
By a theorem of Brouwer (\cite[Theorem 7.12]{mad}),
$\hat \SSS (\mengeW)$ is open in $\setR^{n+1}$ and $\hat \SSS|_\mengeW : \mengeW
\to \hat \SSS (\mengeW)$ is a homeomorphism. Note that with $\hat
\SSS|_\mengeW$, also $\SSS|_{\mengeW \times \setR}$ is a homeomorphism and that
$\SSS (\mengeW \times \setR)$ equals the open set $\hat \SSS (\mengeW) \times
\setR$. We will simply write $\hat \SSS$ and $\SSS$ in place of $\hat
\SSS|_\mengeW$ resp.\ $\SSS|_{\mengeW \times \setR}$. Noting that $\teh = \seh =
(u,x\ee)$, we have $\teh = \seh \to \hat \SSS$ uniformly on compact sets as
$\ve \to 0$, by Proposition \ref{konv fkt}.

Let $p=(0,x_p,v_p)$ be a point of the hyperplane $\uu =0$,
$q:=\SSS(p)=(0,x_q,v_q)$, $\hat p=(0,x_p)$ and $\hat q=\hat \SSS(\hat p)=
(0,x_q)$. Let $R \subseteq \setR^n$ be a bounded open set satisfying
$\overline{R}^{\, \circ} = R$ containing $x_p$. Choose
$\alpha \in (0,\min(a,b))$ and $\lambda>0$ such that $(-a,\alpha] \times
\overline{R_\lambda} \subseteq \mengeW$ where $R_\lambda:=R + \kugel_\lambda
(0)$. Then $\SSS$ is injective on $(-a,\alpha] \times \overline{R_\lambda}
\times \setR$. By property (E), we can assume w.l.o.g.\ (making $\alpha$ smaller
if necessary) that there exists $\ve_1 \in (0,1]$ such that also $(\te)\ee$ and
$(\se)\ee$ are injective on $(-a,\alpha] \times \overline{R_\lambda} \times
\setR$ for all $\ve \le \ve_1$. Defining $\mengeG:= (-a,\alpha) \times R_\lambda
\times \setR$, we have, in particular, that $\SSS$, $\te$ and $\se$ (for $\ve
\le \ve_1$) are injective on $\mengeG$.

Fix $\gamma \in (0,\alpha)$ and $\beta \in [\gamma,a)$. Choose $\delta >0$ with
$\hat \SSS^{-1} (\overline{\kugel_{3 \delta}(\hat q)}) \subseteq (-\beta,\gamma)
\times R$, i.e.\ $\overline{\kugel_{3 \delta}(\hat q)} \subseteq \hat \SSS (
(-\beta,\gamma) \times R)$. Let $\mu \in (\beta,a)$. Choose $\eta \ge \delta$
and $\ve_2 \le \ve_1$ such that
\begin{equation*}
\| \veps - \we \|_{\infty,[-\mu,\alpha] \times \overline{R_\lambda} \times \setR}
= \| h\ee \|_{\infty,[-\mu,\alpha] \times \overline{R_\lambda}} < \eta
\end{equation*}
for all $\ve \le \ve_2$. Since $\SSS (\mengeW \times \setR)= \hat \SSS (\mengeW)
\times \setR$, it follows that $\overline{\kugelz_{3 \delta, 2\eta +
\delta}(q)}=\overline{\kugel_{3 \delta} (\hat q)} \times [v_q-(2 \eta +
\delta),v_q + (2 \eta + \delta)]$ is a compact subset of $\SSS (\mengeW \times
\setR)$. Now let $I$ be a bounded open interval in $\setR$ such that 
$\SSS^{-1} \Big( \overline{\kugelz_{3 \delta, 2\eta + \delta}(q)} \Big)
\subseteq (-\beta,\gamma) \times R \times I
=: \mengeP$ 
which is possible since only the last component of $\SSS$ is dependent on $\vo$ and this dependence is a linear one. Applying $\SSS$ to both sides of this inclusion yields
\begin{equation} \label{inklusion 1}
\overline{\kugelz_{3 \delta, 2\eta + \delta}(q)}
\subseteq \SSS(\mengeP).
\end{equation}
Observe that $p \in \mengeP \subseteq \overline{\mengeP} \cptin \mengeG$ and $q
\in \SSS (\mengeP)$.
Again by Proposition \ref{konv fkt}, there exists $\ve_0 \le \ve_2$ such
that
\begin{equation*}
\| \seh - \hat \SSS \|_{\infty,\overline{\mengeP}} < \frac \delta 2 \quad
\mbox{and} \quad \| \we - \wk \|_{\infty,\overline{\mengeP}} < \frac \delta 2
\end{equation*}
for all $\ve \le \ve_0$. 
The set $\mengeQ_0':= \SSS(\mengeP) \bs (\mypartial \SSS(\mengeP) +
\overline{\kugelz_{\delta,\delta}(0)})$ is open and bounded since $\SSS
(\mengeP)$ has these properties. By \eqref{inklusion 1} and by definition of
$\mengeQ_0'$,
\begin{equation} \label{inklusion 2}
\overline{\kugelz_{2\delta,2\eta}(q)} \subseteq \mengeQ_0'
\end{equation}
holds. Now let $\mengeQ'$ be the connected component of $\mengeQ_0'$ containing
$q$, hence also containing the (connected) set
$\overline{\kugelz_{2\delta,2\eta}(q)}$. Obviously, $\mengeQ'$ is open, bounded
and connected.

Now we apply Theorem \ref{t: penrose-thm allg} for the first time, with
$\mengeG$, $\SSS$, $s_{\ve_0}$, $\mengeQ'$, $q$, $\delta$, $\delta$ and
$M':=\SSS^{-1}(\overline{\mengeQ' + \kugelz_{\delta,\delta}(0)})$ in place of
$U$, $f$, $g$, $W$, $y$, $\delta$, $\eta$ and $A$ to arrive at
\begin{equation*}
\overline{\mengeQ'}
\subseteq s_{\ve_0} (M')^\circ
\subseteq s_{\ve_0} (\overline{\mengeP})^\circ.
\end{equation*}

We now set out to apply Theorem \ref{t: penrose-thm allg} once more to derive an
analogous statement with respect to $\te$. Similarly to above, set $\mengeQ_0 :=
\mengeQ' \bs (\mypartial \mengeQ' + \overline{\kugelz_{\delta,\eta}(0)})$.
Again, $\mengeQ_0$ is open and bounded. By \eqref{inklusion 2} and with
$\mengeQ$ denoting the connected component of $\mengeQ_0$ containing $q$, we
have $\overline{\kugelz_{\delta,\eta}(q)} \subseteq \mengeQ$.

Applying Theorem \ref{t: penrose-thm allg} again, this time with respect
to $\mengeG$, $s_{\ve_0}$, $\te$ (for fixed $\ve \le \ve_0$), $\mengeQ$, $q$,
$\delta$, $\eta$ and $M := s_{\ve_0}^{-1}
(\overline{\mengeQ + \kugelz_{\delta,\eta}(0)})$ in place of $U$, $f$, $g$,
$W$, $y$, $\delta$, $\eta$ and $A$, we obtain
\[
\overline{\mengeQ} \subseteq \te (M)^\circ \subseteq \te
(\overline{\mengeP}).
\]
Hence, $\te$ being a homeomorphism on $\mengeG
$ and $\overline{R}^{\, \circ} = R$, the inclusion $\overline{\mengeQ}
\subseteq \te (\mengeP)$ holds for all $\ve \le \ve_0$.
\end{proof}

\begin{remark} \label{mengendiskussion 1}
An inspection of the preceding proof reveals that $\mengeP$ can be chosen as
having the form $(-\beta,\gamma) \times R \times I$ where $-\beta <0$ is
arbitrarily close to $-a$, the sets $R$ and $I$ are arbitrarily large, yet
bounded open sets ($I$ being of a certain minimum size depending on $\| h\ee
\|_\infty$ on compact sets for small $\ve$) and $\gamma$ has to be sufficiently
small, depending (via $\alpha$) on $R$ and the injectivity behaviour of $\SSS$,
$(\te)\ee$ and $(\se)\ee$ for $\uu>0$.
\end{remark}

If the functions $\te$, $\se$ in Theorem \ref{ich bin die groesste!} are
representatives of generalized functions $\tk$ and $\sk$ then $\tk$ is
invertible around any point on the shock hyperplane, provided $(\te)\ee$
satisfies property (E+):

\begin{theorem} \label{ich bin so toll!}
Let $(\te)\ee$, $(\se)\ee$ and $\SSS$ be as in Theorem \ref{ich bin die groesste!}. If, in addition, $(\te)\ee$ has property (E+) and
\[
\tk:=[(\te)\ee] \, \in \, \calG [(-a,b) \times \setR^n \times \setR,(-a,b) \times \setR^n \times \setR]
\]
and
\[
\sk:=[(\se)\ee] \, \in \, \calG [(-a,b) \times \setR^n \times \setR,(-a,b) \times \setR^n \times \setR],
\]
then, for every $p$ on the hyperplane $\uu=0$, there exists an open neighbourhood $\mengeA$ of $p$ in $(-a,b) \times \setR^n \times \setR$ such that $\tk$ is invertible on $\mengeA$ with inversion data $[\mengeA,\setR^{n+2},\rk,B,\mengeQ]$ where $\rk \in \calG [\setR^{n+2},\mengeD]$ and $B$, $\mengeQ$ and $\mengeD$ are suitable bounded open subsets of $(-a,b) \times \setR^n \times \setR$ with $\mengeQ \subseteq B$ and $\mengeA \subseteq \mengeD$.
\end{theorem}

\begin{proof}
Let $\alpha$, $R_\lambda$, $\mengeG=(-a,\alpha) \times R_\lambda \times \setR$,
$\mengeP$, $\mengeQ$ and $\ve_0$ be as in the proof of Theorem \ref{ich bin die
groesste!}. Recall that under these assumptions, $\te$ is injective on
$\mengeG$ and the inclusions $p \in \mengeP \subseteq \overline{P} \cptin
\mengeG$ and $\overline{\mengeQ} \subseteq \te (\mengeP)$ hold. 
Assuming that $\alpha$ was chosen according to property (E+), there exist $\ve'
\le \ve_0$, $C'>0$ and $N' \in \setN$ such that
\begin{equation} \label{inf bed}
\inf_{(\uu,X,V) \in \mengeG} |\det (\Diff \te (\uu,X,V))| \ge C' \ve^{N'}
\end{equation}
for all $\ve \le \ve^\prime$. Let $\mengeA$ and $\mengeD_1$ be open subsets of
$\mengeG$ such that
$\overline{\mengeP} \cptin \mengeA \subseteq \overline{\mengeA} \cptin \mengeD_1
\subseteq \overline{\mengeD_1} \cptin \mengeG$. Then $p \in \mengeA$ and $K\ee
:= \te (\overline{\mengeA})$ is compact for all $\ve \le \ve_0$. Obviously,
the estimate from below in \eqref{inf bed} is also valid for all $(\uu,X,V) \in
\mengeD_1 \subseteq \mengeG$.

We now apply \cite[Proposition 5.4]{EEinvers} to $(-a,b)
\times \setR^n \times \setR$, $\mengeD_1$, $(\te)\ee$,
$(({\te|_{\mengeD_1}}\kern-.2pt)^{-1})\ee$, $p$, $\{ p \}$, $\overline{\mengeA}$
and
$K\ee$ (in place of $U$, $W$, $(u\ee)\ee$, $(\veps)\ee$, $[(\tilde x\ee)\ee]$,
$K'$, $K$ and $K\ee$ in the notation of \cite{EEinvers}).
Essentially, this (technical) proposition states the following: If a
moderate net $(u\ee)\ee \in \calE_M (U)^n$ with all $u\ee$ injective on a
relatively compact open subset $W$ of $U$ satisfies an estimate corresponding to
\eqref{inf bed} on $W$, then the inverses $v\ee$ of $u\ee|_W$ can be extended to
a uniformly bounded moderate net $(\tilde v\ee)\ee \in \calE_M (\setR^n)^n$ in
such a way that $\tilde v\ee|_{u\ee(K)} = v\ee|_{u\ee(K)}$ and $\tilde
v\ee(x)=y_0$ on $\setR^n \bs u\ee(W)$, where $y_0 \in \setR^n$ and the compact
subset $K$ of $W$ can be arbitrarliy prescribed. Therefore,
there exist extensions $\re$ of $({\te|_{\mengeD_1}})^{-1}$ with $\re|_{K\ee} =
(({\te|_{\mengeD_1}})^{-1})|_{K\ee}$ and $\re (x)=p$ on $\setR^{n+2} \bs \te
(\mengeD_1)$ such that $(\re)\ee \in \calE_M (\setR^{n+2})^{n+2}$. In
particular, the proposition ensures that the
net $(\re)\ee$ is c-bounded into any
(bounded) open subset $\mengeD$ of $\setR^{n+2}$ that contains the convex hull
of $\overline{\mengeD_1} \cup \{ p \} = \overline{\mengeD_1}$. As to the
last statment, see the proof of \cite[Proposition 5.3\,(2)]{EEinvers} where
$\tilde v\ee$ (i.e. $\re$, in the case at hand) is explicitly constructed.

Set $\rk := [(\re)\ee] \in \calG [\setR^{n+2},\mengeD]$. On the one hand,
 we have $\mengeQ \subseteq \te (\mengeP) \subseteq
\te(\mengeA) \subseteq K\ee$ and, therefore, $\re (\mengeQ) =
({\te|_{\mengeD_1}})^{-1} (\mengeQ) \subseteq \mengeP \subseteq
\overline{\mengeP} \cptin \mengeA$, implying that $(\re|_\mengeQ)\ee$ is
c-bounded into $\mengeA$. Moreover,
\[
\te \circ \re |_\mengeQ = \te \circ \te^{-1}|_\mengeQ = \id_\mengeQ,
\]
establishing $[\mengeA,\setR^{n+2},\rk,\mengeQ]$ as a right inverse of $\tk$ on $\mengeA$. On the other hand, since $\te (\mengeA) \subseteq K\ee$, we have
\[
\re \circ \te |_\mengeA = \re|_{K\ee} \circ \te |_\mengeA = \te^{-1}|_{K\ee} \circ \te |_\mengeA =\id_\mengeA.
\]
By c-boundedness of $(\te)\ee$, there exists $K' \cptin (-a,b)
\times \setR^n \times \setR$ with $\te(\overline{\mengeA}) \subseteq K'$ for
sufficiently small $\ve$. Hence, $(\te|_\mengeA)\ee$ is c-bounded into any
(bounded) open set $B$ containing $K'$. It follows that
$[\mengeA,\setR^{n+2},\rk,B]$ is a left inverse of $\tk$ on $\mengeA$. Combining
these results, we obtain that $\tk$ is invertible on $\mengeA$ with inversion
data $[\mengeA,\setR^{n+2},\rk,B,\mengeQ]$.
\end{proof}

Having collected the necessary tools, we can now establish the main result on
invertibility of the generalized coordinate transformation $\tk$.

\begin{theorem}\label{main result}
The generalized coordinate transformation $\tk=[(\te)\ee]$ is locally invertible
(in the sense of Definition \ref{d: local invertibility}) on some open set
$\Omega$ containing the half space $(-\infty,0] \times \setR^3$.
\end{theorem}

\begin{proof}
By Proposition \ref{falsch}, $(\te)\ee$ as well as $(\se)\ee$ possess property
(E+). More\-over, $\hat \SSS$ is injective on some open set $\mengeW$ containing
$(-\infty,0] \times \setR^2$ by Lemma \ref{schlange}. Then, by Theorem \ref{ich
bin so toll!}, for every $p$ on the hyperplane $\uu =0$ there exists an open
neighbourhood $\mengeA (p) \subseteq \setR^4$ such that $\tk$ is invertible on
$\mengeA (p)$. Recall that each $\mengeA (p)$ contains some set $\mengeP =
(-\beta,\gamma) \times R \times I$ as discussed in Remark \ref{mengendiskussion
1}. In particular, all of $\beta>0$, $R$ and $I$ (both bounded) can be chosen
arbitrarily large. Forming the union $\Omega$ of a family of $\mengeA (p)$ with
the corresponding sets $\mengeP$ covering the left half space, we obtain that
the generalized function $\tk$ is locally invertible on $\Omega$, constituting
an open set containing $(-\infty,0] \times \setR^3$.
\end{proof}

\section*{Acknowledgements}
The authors were supported by START-project Y-237 and by
project P20525 of the Austrian Science Fund.

%\bibliographystyle{gAPA}
%%%%%\bibliographystyle{plain}
%%%%%\bibliography{erlacher_grosser_gf2009_arxive_100319}{}

\end{document}